# Physical Analysis of Bennu Samples Reveals Regolith Production by Collisional Disruption on Near-Earth Asteroids

R.-L. Ballouz[1*], A.J. Ryan[2], R.J. Macke[3], O.S. Barnouin[1], M. Lê[4], J. Moreno[4], S. Eckley[5], L. Hanton[6], A. Hildebrand[6], V. Toy-Edens[1], R.M. Meier[1], M. Berkson[1], E. Asphaug[2], S. Cambioni[7], C.G. Hoover[8], K. Jardine[8], E.R. Jawin[9], N. Lunning[10], J.L. Molaro[11], M. Pajola[12], K. Righter[13], K.T. Ramesh[1,4], F. Tusberti[12], K.J. Walsh[14], C.W.V. Wolner[2], D.N. DellaGiustina[2], H.C. Connolly, Jr.[2,15,16], D.S. Lauretta[2].

[1]*Johns Hopkins University Applied Physics Laboratory, Laurel, MD, USA*
[2]*Lunar and Planetary Laboratory, University of Arizona, Tucson, AZ, USA*
[3]*Vatican Observatory, Vatican City State*
[4]*Hopkins Extreme Materials Institute, Johns Hopkins University, Baltimore, MD, USA*
[5]*Jacobs-JETS, NASA Johnson Space Center, Houston, TX, USA*
[6]*University of Calgary, Calgary, Canada*
[7]*EAPS, Massachusetts Institute of Technology, Cambridge, MA, USA*
[8]*Arizona State University, Tempe, AZ, USA*
[9]*National Air and Space Museum, Smithsonian Institution, Washington, DC, USA*
[10]*Astromaterials Research and Exploration Science (ARES), NASA Johnson Space Center, Houston, TX, USA*
[11]*Planetary Science Institute, Tucson, AZ, USA*
[12]*INAF, Astronomical Observatory of Padova, Padova, Italy*
[13]*University of Rochester, Rochester, NY, USA*
[14]*Southwest Research Institute, Boulder, CO, USA*
[15]*Department of Geology, Rowan University, Glassboro, NJ, USA*
[16]*Department of Earth and Planetary Sciences, American Museum of Natural History, New York, NY, USA*
*ronald.ballouz@jhuapl.edu

**Abstract**
Owing to the extremely low gravity of small near-Earth asteroids (NEAs), it has been assumed that impact-generated rock fragments escape into space and thus do not contribute to the accumulation of regolith. However, centimeter-sized stones returned from the small NEA Bennu by NASA's OSIRIS-REx mission exhibit impact craters up to a few millimeters wide, implying that impact fragments and impact-processed rocks are retained despite the microgravity environment. To understand how, we combined detailed physical analysis of Bennu samples, laboratory experiments of impacts into simulant rocks, and 3D numerical simulations of disruptive impacts into boulders. We find that the majority (~85% by mass) of impact fragments eject toward and penetrate the asteroid's weak, porous surface, leading to their retention. In addition, crater depth-to-diameter ratios ($d/D$) suggest that the Bennu samples (median crater $d/D = 0.36 \pm 0.1$) are structurally representative of the asteroid's large boulders (median crater $d/D = 0.33 \pm 0.08$, measured previously). Our analyses indicate that most of Bennu's surface rocks (those with diameters ≲ 20 m) could be products of in situ collisional disruption. This impact-driven mechanism of regolith production likely occurs on other small NEAs with highly porous surfaces.

# 1. Introduction

Knowledge of the origin, evolution, and physical properties of regolith on near-Earth asteroids (NEAs) is important for understanding the geologic context of returned samples [Tsuchiyama et al. 2011, Nakamura et al. 2023, Lauretta & Connolly et al. 2024, Connolly & Lauretta et al. 2025], developing inferences about the poorly understood interiors of asteroids [DellaGiustina & Ballouz et al. 2024, Ballouz et al. 2025], and formulating planetary defense strategies for mitigating impact hazards to Earth [Stickle et al. 2017, NEOWARP 2025].

High-resolution spacecraft images (<1 m/pixel) of small (kilometer-scale) NEAs such as the stony asteroids (25143) Itokawa [Fujiwara et al. 2006], (65803) Didymos and its satellite Dimorphos [Barnouin et al. 2024], as well as the carbonaceous asteroids (162173) Ryugu [Sugita et al. 2019] and (101955) Bennu [Lauretta & DellaGiustina et al. 2019a, DellaGiustina & Emery et al. 2019], revealed surfaces dominated by boulders (Fig. 1). The shared trait of boulder-rich surfaces on such asteroids, regardless of their different compositions, suggests a rubble-pile structure [Walsh 2008, Pajola et al., 2024, Ballouz et al. 2025] for their interiors and a common evolution for their surfaces (Fig. 1). The largest NEAs, such as Eros (Fig. 1c), exhibit a lower surface density of boulders (see Fig. 7 of Pajola et al. [2024]), which may point to a significant change in either regolith production mechanisms or surface age with asteroid size. A detailed understanding of regolith production on Bennu, enabled by the samples returned by the OSIRIS-REx mission [Lauretta & Connolly et al. 2024], may provide broad insight into small asteroid surface evolution.

*1.1 The Origin of Asteroidal Regolith: Insights from Remote Sensing*
The collisional lifetimes of kilometer-scale NEAs are much shorter than the age of the Solar System, so they are inferred to be the products of the catastrophic disruption of larger parent bodies and have likely experienced multiple subsequent disruption events [Walsh et al. 2024]. In the case of asteroid binaries, satellites such as Dimorphos are thought to form through rotational disruption of the primary asteroid [e.g., Barnouin et al. 2024, Walsh 2018, Pajola et al., 2024]. As such, the boulder-rich regolith of these small NEAs may be inherited from their disrupted parent bodies [Barnouin-Jha et al. 2008].

For larger airless bodies, from the Moon to >10-km-diameter asteroids, meteoroid impacts are thought to contribute to the development of a layer of fine particulate regolith, as their surface gravity is sufficiently large to retain crater ejecta. In contrast, for kilometer-scale asteroids, meteoroid impacts were historically thought to result in gradual erosion, owing to the low surface acceleration that would allow fine ejecta to escape into space [Housen et al. 1979].

However, despite the boulder-rich appearance of small NEAs, spacecraft observations have demonstrated that fine regolith is present on the surface, albeit in scarce amounts for some asteroids such as Bennu [e.g., Walsh et al. 2019]. This spurred the development of a non–impact-driven hypothesis for the production of fines: thermal cycling leading to particle fracturing [Delbo et al. 2014; Molaro et al. 2020a,b]. OSIRIS-REx afforded insights into these various surface processes through the thermal and physical characterization of Bennu's carbonaceous surface material at decimeter to decameter scales [e.g., Rozitis et al. 2020, Ballouz et al. 2020, Jawin et al. 2023, Cambioni et al. 2021].

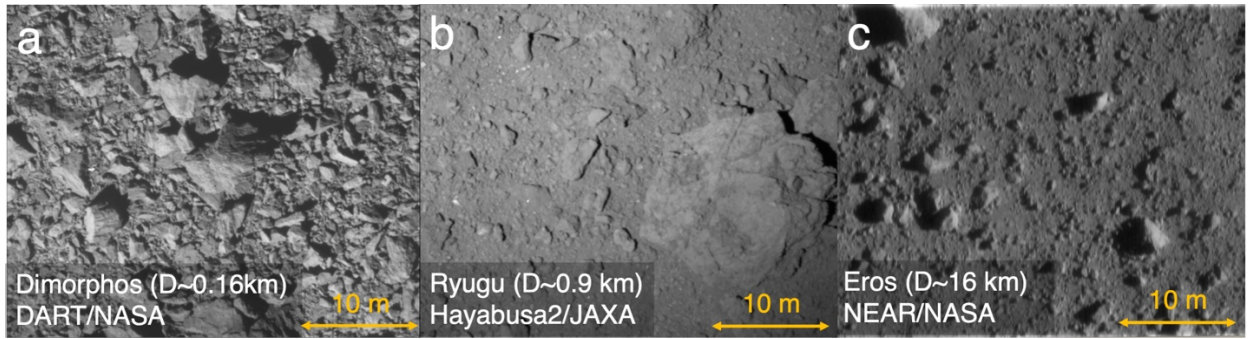

**Figure 1.** Same-scale images of the bouldery surfaces of **a,** stony asteroid Dimorphos, which has a diameter, $D \sim 0.16$ km; **b,** the carbonaceous asteroid Ryugu, $D \sim 0.9$ km; and **c,** the stony asteroid Eros, $D \sim 16$ km. The boulder density on the surfaces of asteroids is heterogenous; however, larger asteroids tend to have a lower number of large boulders per unit area.

Analysis of Bennu's surface showed that boulders can effectively armor against smaller impacts, inhibiting the formation of ≲2 m craters in the regolith [Bierhaus et al. 2022]; instead, small craters form on boulders [Ballouz et al. 2020]. Ejecta from impacts into boulders, in combination with fracturing by thermal fatigue, have been proposed to explain the sporadic ejection of particles from Bennu [Lauretta & Hergenrother et al. 2019, Bottke et al. 2020, Molaro et al. 2020b], the majority of which appear to fall back to the surface [Chesley et al. 2020]

In addition, regolith production on Bennu has been hypothesized to be influenced by the heterogeneity of its boulders, which comprise two broad populations with distinct reflectance, thermal, and physical properties [DellaGiustina et al. 2020, Rozitis et al. 2020, Jawin et al. 2023] that may break down differently [Cambioni et al. 2021]. Brighter, smoother, more angular boulders are hypothesized, based on their higher thermal inertias, to be denser and stronger than the darker, rougher, hummocky boulders, which are more prevalent [Rozitis et al. 2020, DellaGiustina et al. 2020]. Cambioni et al. [2021] proposed that the darker, more porous boulders on Bennu may preferentially compact rather than fragment (as previously suggested for the largest craters on the 60-km-diameter carbonaceous asteroid (253) Mathilde [Housen et al. 1999]), whereas the brighter, less porous boulders more easily comminute into smaller fragments. In this model, the scarcity of fines on Bennu's surface is driven by inhibited production due to the predominance of the darker boulder population [Rozitis et al. 2020, DellaGiustina et al. 2020, DellaGiustina & Emery et al. 2019]. Alternatively, the uppermost layer of regolith may be of sufficiently high porosity (>50%), as indicated by its response to the OSIRIS-REx sampling event [Walsh & Ballouz et al. 2022, Lauretta et al. 2022], that any fines produced by boulder comminution would percolate to deeper layers [Bierhaus et al. 2023].

*1.2 The Limits of Collisionally Driven Regolith Production on Small NEAs*

A central problem with assessing the feasibility of these hypotheses is the poorly understood strength properties of asteroid surface materials. Understanding surface strength also has important implications for interpreting the cratering record [e.g., Bottke et al. 2020, Bierhaus et al. 2022] and understanding the outcomes of kinetic impactor mitigation of potentially hazardous asteroids [e.g., Raducan et al. 2022]. Impact craters on small bodies form in one of two primary regimes: the *strength* regime, where the material's cohesive and tensile properties dominate and control crater size, and the *gravity* regime, where the target's gravitational acceleration is the limiting factor for crater growth. On weak, low-gravity bodies, craters can form in either regime [Ballouz et al. 2020, Arakawa et al. 2020], complicating interpretations of surface age and mechanical properties.

For carbonaceous asteroids, this issue is exacerbated by biases in the meteoritic record. Carbonaceous chondrites are scarce on Earth, probably because passage through the atmosphere, impact, and exposure to terrestrial weathering lead to the preferential disintegration and destruction of these comparatively weak materials [e.g., Shober et al. 2025]. Thus, destructive strength measurements using representative materials are rare.

Nevertheless, spacecraft missions have provided key insight. The Small Carry-on Impactor (SCI) fired by the Hayabusa2 mission into Ryugu's regolith formed a crater in the gravity regime [Arakawa et al. 2020], indicating a surface with little strength (1.3 Pa). Commensurate with these findings, direct and remote sensing measurements by the OSIRIS-REx mission showed that the surface cohesion of the regolith on Bennu is extremely low (≤2 Pa ) [Lauretta et al. 2022, Walsh & Ballouz et al. 2022, Perry et al. 2022]. In contrast, Bennu's boulders are estimated to be much stronger than the surface regolith, on the order of 1 MPa [Ballouz et al. 2020], indicating that crater formation on these rocks should be governed by their material properties rather than gravity.

Following an impact, the fraction of ejected mass that is retained, $M(v<v_{esc})$, can be estimated by considering cratering scaling relationships for the distribution of ejecta speeds and comparing those to the escape speed, $v_{esc}$:

$$M(v < v_{esc}) = 1 - (c_s v_{esc} \sqrt{\left(\frac{\rho}{Y}\right)})^{-\beta}, \qquad (1)$$

where $c_s$ and $\beta$ are material scaling constants, $\rho$ is the target density, and $Y$ is the target strength [Nakamura 2017]. Figure 2 shows how the fraction of retained ejecta mass scales with $Y$ for different cases of $v_{esc}$, using Equation 1.

Some material is retained in the case of impacts into NEA regolith, because a fraction of the ejecta has speeds below the escape velocity (generally $v_{esc}$ ≲1 m/s). Low-speed ejecta were directly observed during the SCI experiment, demonstrating that even energetic impact events can result in localized fallback, rather than widespread loss of fragments [Arakawa et al. 2020]. Similarly, on Bennu, ejecta deposits associated with Bralgah crater [Perry et al. 2022] and the crater formed by the OSIRIS-REx sampling event [Lauretta et al. 2022] show that a substantial portion of the impact-mobilized regolith remained gravitationally bound.

However, these types of impacts are expected to generate minimal (compared to the volume of ejected material) new regolith through rock fragmentation, as the shear stress required to excavate most of a gravity-scaled crater's volume (1.3 Pa) [Arakawa et al. 2020] is much smaller than the strength of the constituent material (~1 MPa) [Ballouz et al. 2020]. We assume that Ryugu material has a similar impact strength as that of Bennu based on their similar thermal characteristics based on remote sensing [Grott et al. 2019, Rozitis et al. 2020]. In comparison, cratering impacts into boulders on Bennu and Ryugu are expected to produce high-velocity ejecta that exceed the escape speed of the host asteroid (Fig. 2), resulting in their permanent loss.

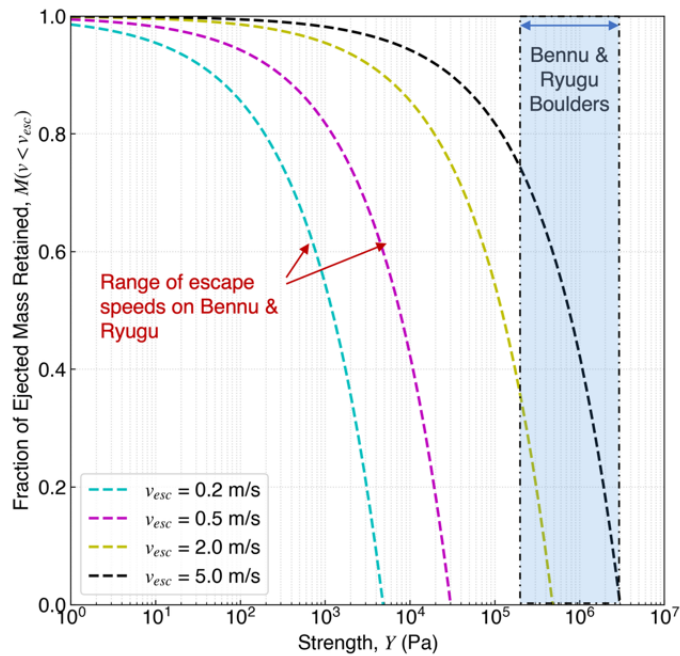

**Figure 2.** Fraction of ejected mass retained, $M(v<v_{esc})$, from cratering impacts into material with strength $Y$, and assuming material constants similar to that of weakly cemented basalt, $c_s = 0.122$ and $\beta = 1$ [Nakamura 2017], for $\rho = 1.8$ g/cm³. The cyan, magenta, yellow, and black dashed curves show $M(v<v_{esc})$ for $v_{esc}$ = 0.2, 0.5, 2.0, and 5.0 m/s. The escape speeds for Bennu and Ryugu range from 0.2 to 0.5 m/s. Cratering impacts into Bennu and Ryugu boulders, which have an estimated $Y \sim 0.2$–2 MPa [Ballouz et al. 2020], would lead to all of the fragments escaping the asteroid.

*1.3 Sample Analysis as a Window into Asteroid Regolith Development*
The mechanism for regolith production and loss has implications for the expected residence times of particles on the surface of NEAs, which is necessary for providing geologic context for the returned samples and meteorites [Lauretta & Connolly et al. 2024, Connolly & Lauretta et al. 2025, Nakamura et al. 2022]. Processes such as space weathering and cosmogenic exposure, recorded in noble gas and nitrogen isotopic signatures, are time-dependent and sensitive to both the depth and duration of surface residence. For example, galactic cosmic rays produce isotopic signatures that constrain exposure durations at meter-scale shielding depths. Surface and subsurface samples returned from Ryugu by Hayabusa2 indicate cosmic ray exposure (CRE) ages of about 5 Myr [Okazaki et al. 2022] — much older than the <1 Myr age predicted for the upper 2–4 m of regolith on Ryugu based on gravity scaling and crater statistics [Takaki et al. 2022]. This discrepancy suggests that surface materials on small NEAs may be retained for longer than crater scaling relationships would imply.

Bennu samples exhibit CRE durations ranging mostly from 1 to 3 million years [Marty et al. 2025], consistent with multiple independent constraints: (i) surface exposure ages of 2–7 Ma from radionuclide systematics [Keller & Thompson et al. 2025]; (ii) small crater retention ages of 1.6–2.2 Ma based on crater statistics [Bierhaus et al. 2022]; and (iii) an estimated dynamical lifetime of 1.75 ± 0.75 Ma since Bennu became decoupled from the main asteroid belt (Ballouz et al. 2020).

In contrast, solar wind–derived noble gases, which implant only into the outermost microns of grain surfaces, constrain much shorter exposure durations (1–3 Myr), though these are more sensitive to regolith disturbance and burial processes [Marty et al. 2025]. Accurate reconstruction of these histories and the significance of these measurements requires a mechanistic understanding of how regolith is produced, mobilized, buried, and lost in the low-gravity environments of NEAs.

The physical and thermal properties of Bennu samples [Ryan et al. in press] provide vital clues to the development of regolith on NEAs, particularly when placed in the context of remote observations. Analysis of Bennu samples has revealed at least two distinct lithologies, termed hummocky and angular based on their distinctive particle morphologies [Lauretta & Connolly et al. 2024, Connolly & Lauretta et al. 2025], with visible similarities to Bennu's boulders [DellaGiustina et al. 2020, Jawin et al. 2023] despite the difference in scale [Ryan et al. in press]. The hummocky lithology has high surface roughness, relatively low density, and similar morphology to the darker, lower–thermal-inertia boulder population. The angular lithology has higher density and resembles the brighter, higher–thermal-inertia boulders. The distinct densities of these two lithologies validate inferences from remote sensing about the differing porosity and strength of the corresponding boulder populations [e.g., Rozitis et al. 2020, Cambioni et al. 2021, Jawin et al. 2023]. Also observed in the samples are mottled particles, which have been posited to relate to vein-bearing boulders on Bennu [Lauretta & Connolly et al. 2024, Jawin et al. 2023, Kaplan et al. 2020], as well as particles that cannot be classified as angular, hummocky, or mottled [Connolly & Lauretta et al. 2025].

Here, we combine physical analysis of Bennu samples with experiments and simulations of hypervelocity impacts to shed new light into regolith development on this and other small NEAs. In Section 2, we describe craters of up to a few millimeters on centimeter-scale Bennu particles. In Section 3, we present laboratory experiments of hypervelocity impacts into a simulant material. In Section 4, we use the results of these experiments to calibrate numerical models of hypervelocity impacts into boulders to investigate the dynamics of the resultant fragments. In Section 5, we use video-documentation of sample curation activities to estimate the angle of friction and evaluate the penetration dynamics of impact-generated fragments into the surface of Bennu. Finally, in Section 6, we calculate surface residence times of centimeter-scale fragments and compare them to the exposure ages of Bennu samples determined in other studies. Based on our findings, we present a mechanism for production and retention of regolith on small NEAs.

## 2. Craters on Bennu Samples

A set of hummocky and angular Bennu stones were analyzed using structured light scanning (SLS) and X-ray computed tomography (XCT) [Lauretta & Connolly et al. 2024; Ryan et al. in press]. These data were used to construct detailed 3D shapes at resolutions of a few to tens of microns. We analyzed the surfaces of 29 stones that range in longest dimension from 0.5 to 3.1 cm (Table A1), using the Dragonfly 3D World software for XCT data [Dragonfly 3D World 2024] and the Small Body Mapping Tool (SBMT) [Ernst et al. 2018] for XCT- and SLS-derived shape models. Measurement uncertainties in this section, and throughout the paper, are reported to 1-$\sigma$.

Using SBMT, we identified and measured a total of 143 craters, with diameters ranging from ~0.05 to 2 mm, on 15 of the stones: seven angular, three hummocky, two mottled, and three unclassified. Examples are shown on XCT-derived shape models in Fig. 3, and detailed shape model and microscopy views are provided for two craters in Fig. 4. The craters were identified based on the appearance of a circular depression on an otherwise flat surface of the stones. These craters are sometimes accompanied by fractures or extended spallation features, though this is not always the case. An additional seven stones had surface depressions, but we could not confidently assess an impact origin. The remaining seven stones showed no indications of craters or surface depressions, despite having similar morphologies to the other stones.

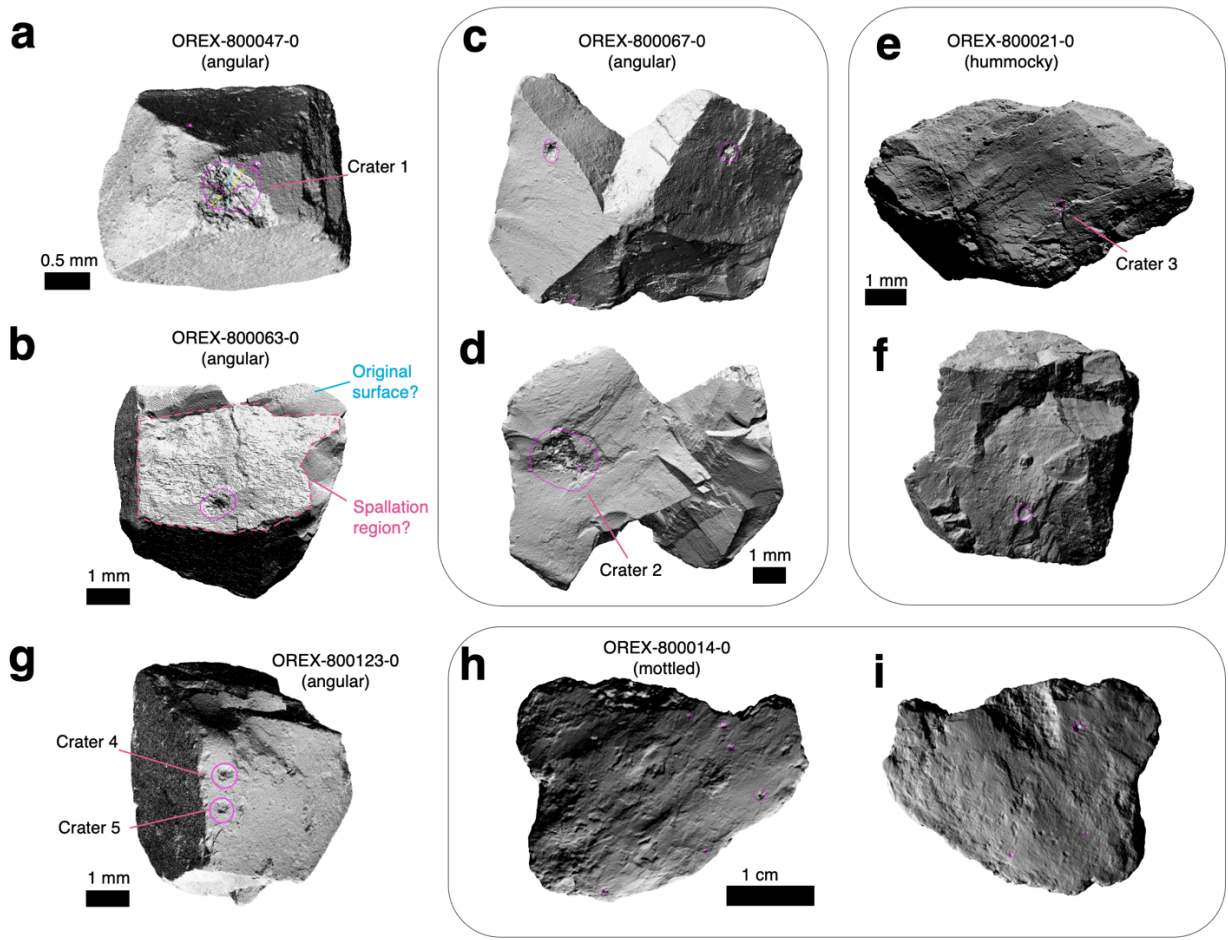

**Figure 3.** Examples of craters (circled in magenta) on Bennu stones. Shown are XCT-derived shape models, visualized in the SBMT. **a,** Crater 1 (~0.6 mm diameter) on angular stone OREX-800047-0. **b,** A crater (~0.8 mm diameter) with a central pit and spallation zone on angular stone OREX-800063-0. The crater occurs on a smooth, flat region of the stone that may itself be an extended spallation region (magenta polygon) formed by the same impact. A hypothesized pre-impact surface is indicated in cyan. **c,** Two craters on angular stone OREX-800067-0. **d**, Crater 2, the largest crater we observed among the samples (2 mm diameter), on the other side of OREX-800067-0. **e,** Crater 3, a sub-millimeter crater that is coincident with a linear fracture on hummocky stone OREX-800021-0. **f,** A crater on the other side of OREX-800021-0. **g,** Craters 4 and 5 (both ~0.5 mm diameter) on angular stone OREX-800123-0. **h**,i, Multiple millimeter-scale craters scattered on both sides of mottled stone OREX-800014-0, the largest stone returned by OSIRIS-REx.

For most stones, multiple faces are cratered. For example, the two faces of the mottled, platy stone OREX-800014-0 have similar crater densities (Fig. 3). These observations imply that the asteroid's near-surface has experienced regolith gardening and/or mass movement, exposing various sides of the regolith particles to meteoroid impacts. This is consistent with remote sensing–based inferences of recent mass wasting on Bennu globally [Jawin et al. 2020] and near the OSIRIS-REx sampling site [Barnouin et al. 2022].

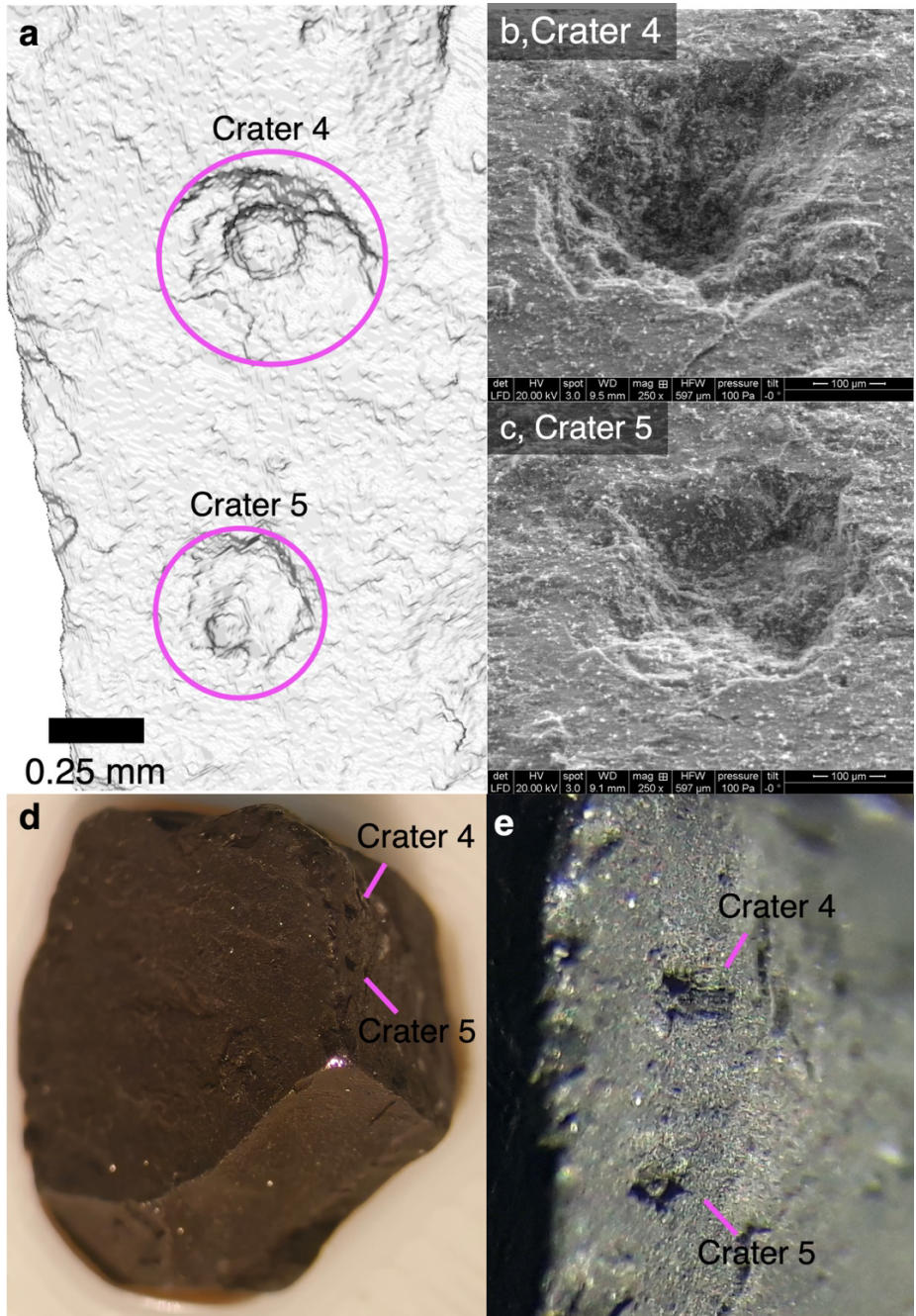

**Figure 4.** Detailed views of craters 4 and 5 (Fig. 3g) on angular stone OREX-800123-0. **a,** Close-up view of the craters in an XCT-derived shape model. **b,c,** Scanning electron microscopy (SEM)

image of craters 4 and 5, respectively. **d,** Microscope context image of OREX-800123-0 with the craters in oblique view. **e,** magnified microscope image of OREX-800123-0 with craters in an oblique view.

In the XCT scans, we observe that some of the larger candidate impact craters exhibit radial fractures that propagate along the surface and into the interior of the particle (Fig. 5b,f). Fig. 5b shows an example of a ~500 µm crater (highlighted in Fig. 3a) that has enhanced X-ray attenuation near its floor, which may indicate impact-induced compaction or melt. Scanning electron microscopy of Bennu millimeter-sized surface particles collected by the OSIRIS-REx contact pads revealed the presence of ~10 µm impacts pits with melt deposits [Keller & Thompson et al. 2025]. In Section 3, we show, through impact experiments, that compaction of phyllosilicates may also be an outcome of impacts on to Bennu material. Regardless, these observations, in sum, bolster confidence in an impact origin for these features. Furthermore, the presence of these features provides further evidence that the OSIRIS-REx mission was able to preserve the physical context of Bennu regolith from the moment of sampling through transport to the NASA JSC curation facility.

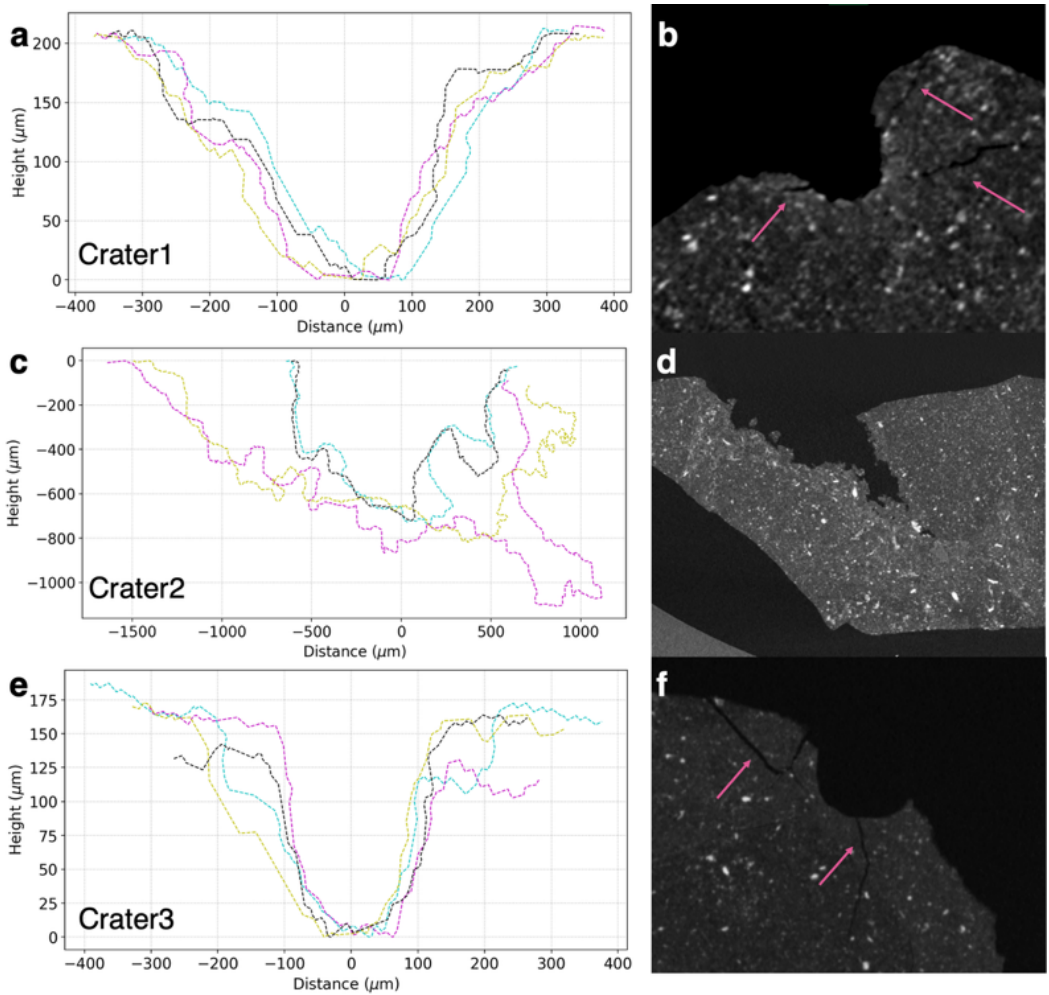

**Figure 5.** Morphological and XCT (~5–10 µm/voxel) details of prominent craters on centimeter-scale Bennu samples. **a,** Four profiles through crater 1 (Fig. 3a) on OREX-800047-0. **b,** XCT slice

of crater 1, showing radial fractures (magenta arrows) and increased X-ray attenuation near the crater floor. **c,** Four profiles of crater 2 (Fig. 3d) on OREX-800067-0. The bulbous end of the crater is most apparent in the magenta profile. **d,** XCT slice of crater 2. **e,** Four profiles of crater 3 (Fig. 3e) on OREX-800027-0. **f,** XCT slice of crater 3, showing linear fractures (magenta arrows) from the crater in its vicinity.

A subset of 40 craters were sufficiently resolved in the XCT-derived shape models that their depths could be accurately measured. As all craters have some natural asymmetry, we mapped four profiles across each of these and measured the depths and diameters from each profile; examples are shown together with XCT slices in Fig. 5. These profiles reveal a paraboloid depression (central pit) and a distinct shallow depression extending outward from the crater interior. We interpret the central pit as the excavation cavity of the impact and the surrounding shallow depression as the spallation region, formed by near-surface tensile failure during crater growth. These features are characteristic of impacts into rocks, ice, and other brittle materials [e.g., Holsapple & Housen 2022].

Moreover, our laboratory experiments of hypervelocity impacts into simulant material (described in Section 3) produced craters with the same diagnostic morphologies — central pits surrounded by shallow spallation depressions. These experiments also revealed associated features such as radial fracturing and localized pore compaction, which match SEM and XCT observations (Figs. 4 and 5b) of the natural craters. The agreement between sample observations and experimental outcomes confirms that the craters preserved on Bennu stones formed by impact, rather than by alternative processes such as desiccation cracking or thermal fatigue.

The size distribution of all craters mapped in this study is shown in Fig. 6a. For the subset of the 40 largest craters whose morphologies could be measured, the diameters (including the spallation region) and depths (from the lowest point on the pit to the edge of the spallation region) are shown in Fig. 6b. The distribution of crater depth-to-diameter ratios ($d/D$) for those 40 craters is shown in Fig. 6c. The median $d/D$ is $0.36 \pm 0.1$, with angular stones having slightly shallower craters ($d/D = 0.35 \pm 0.01$) than hummocky stones ($d/D = 0.39 \pm 0.07$). However, the distributions overlap to within 1-$\sigma$, so it is not immediately clear that the population characteristics are distinct enough to draw conclusions about how crater properties may reflect physical distinctions between these two lithologies [Ryan et al. in press] without further analysis. The distribution of crater $d/D$ values overlaps with that found for decameter-scale boulders on Bennu (median $d/D = 0.33 \pm 0.08$) [Ballouz et al. 2020]. The ranges of $d/D$ values for craters on both the samples and the boulders are wider than those of large craters on the surface of Bennu and other NEAs [Daly et al. 2022, Ballouz et al. 2025], which are typically ~0.1–0.15, though they can be as high as ~0.25 for craters that may be relatively fresh. In addition, the craters on Bennu samples and boulders are deep relative to those in other coherent targets; for instance, the $d/D$ for craters on sandstones is ~0.2 [Poelchau et al. 2013], which is also typical of craters on the Moon and non-porous rock targets [e.g., Nakamura 2017].

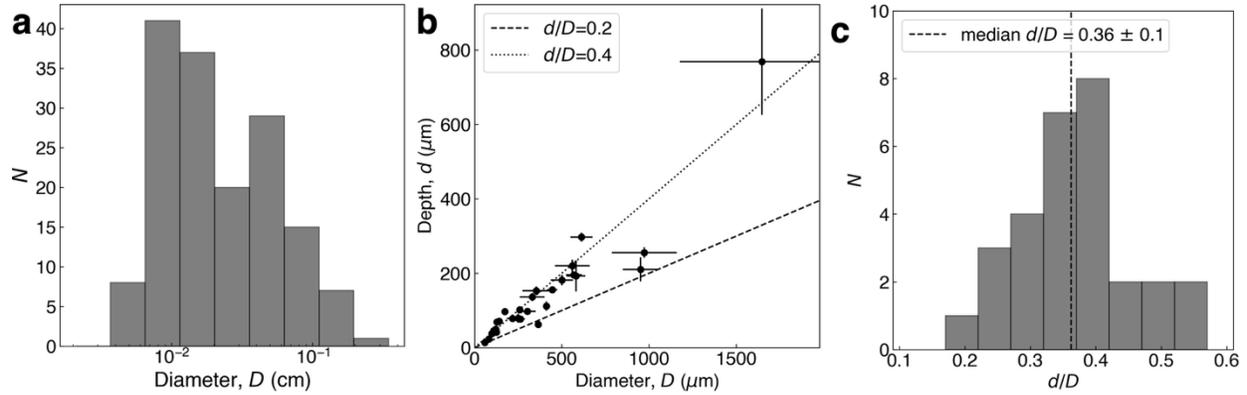

**Figure 6.** Depth, *d*, and diameter, *D*, of the craters on centimeter-scale Bennu samples. **a,** Histogram showing the distribution of all 143 craters mapped on Bennu samples in this study. **b,** The *d* of a subset of 40 of the largest craters as a function of their *D*, showing how the population clusters around *d/D* = 0.4 (dotted black line), with a minority reaching values as small as *d/D* = 0.2 (dashed black line). **c,** A histogram of the crater *d/D* values from panel b, which have a median of 0.36 ± 0.1. The values are not normally distributed, and there are some outlier craters with deep *d/D*.

The relatively large median *d/D* of craters on Bennu's boulders compared to its surface is hypothesized to be due to the porous nature of the former [Cambioni et al. 2021, Delbo et al. 2022], which may allow for compaction. Impacts into terrestrial rocks (e.g., gypsum, pumice, and sandstones) tend to result in deeper craters when targets are more porous [e.g., Nakamura 2017], though other factors such as impactor density and speed may also play a role. Experiments have shown that denser impactors may create deeper craters [e.g., Okamoto et al. 2015]. Direct and indirect analysis of microporosity through ideal gas pycnometry, SLS, and X-ray diffraction (XRD) has shown that angular and hummocky Bennu samples are porous, with values, ranging between 37 and 45% [Ryan et al. in press] —within the ranges estimated for Bennu boulders based on their thermal inertia (~49–55% and ~24–38% for the darker hummocky and brighter angular boulders, respectively) [Rozitis et al. 2020].

The overlap in *d/D* values of craters on Bennu samples and boulders also suggests structural similarity between them, as they appear to respond in a self-similar way to impacts. Thus, the porosities observed in the samples may be reflective of the total porosities of Bennu's boulders, where craters with *d/D* > 0.3 were observed on boulders as large as 10 m in diameter [Ballouz et al. 2020]. The structural properties of the interiors (decimeter to meter scales) of decameter-scale boulders may be similar to their surfaces, which can be probed through infrared measurements of diurnal temperature fluctuations that probe the upper few millimeters of boulders [e.g., Rozitis et al. 2020, Ryan et al. in press].

Not all craters observed on the Bennu samples have the central pit and spall features. Crater 2, on OREX-800067-0 (Fig. 3d), has an irregular morphology reminiscent of craters typically made from highly oblique impacts. It has no clear spall region and exhibits a triangular pyramidal shape with a bulbous pit (Fig. 5c,d). Triangular pyramidal crater shapes have been reported for impacts into porous targets [Nakamura 2017] such as Seeberger sandstones, cement mortar, and weakly cemented basalt; however, these typically have *d/D* < 0.3. In contrast, crater 2 has *d/D* ~ 0.5. This relatively large value may be due to an impact from a denser object, as experiments have shown that the penetration depth of an impactor varies with the impactor-to-target density ratio [Okamoto et al. 2013].

To obtain further insight into how the observed impact features formed, and what they may tell us about impact processes on the surface of Bennu in general, we conducted a series of laboratory and numerical impact experiments, using insights from the physical and thermal characterization of the Bennu samples.

## 3. Laboratory Experiments of Hypervelocity Impacts into Simulant

*3.1 Experimental Setup and Target Characterization with micro-XCT*

We conducted hypervelocity impact experiments into a material simulating CI carbonaceous chondrites (Space Resources Technologies), which thus far appear to be the closest meteoritic analog to Bennu samples [e.g., Lauretta & Connolly et al. 2024; Zega & McCoy et al. 2025]. The chemical and physical properties of the simulant are described in detail by [Metzger et al. 2019]. The simulant has an unconfined compressive strength of approximately 1.7 MPa [Metzger et al. 2019], similar to the impact strength estimated for meter-sized Bennu boulders [Ballouz et al. 2020]. For these experiments, we used $9 \times 9 \times 9$ cm$^3$ cubes.

Before each experiment, we performed micro-X-ray computed tomography (micro-XCT) of the targets to measure their volumes, assess their interior structure and inter-heterogeneity, and to perform before- and after-impact comparisons to assess the effect of hypervelocity impacts on their structural properties (Fig. 8). The micro-XCT scans were taken at the Johns Hopkins University Applied Physics Laboratory's Material Characterization Facility using a North Star Imaging X-50 system with voxel sizes of 51.4–54.0 μm. These data allow for comparisons with the craters we characterized in previous section, although those were captured at smaller voxel sizes by a factor of ~3.

Fig. 8 shows the interiors of the four targets used in the impact experiments. Overall, the appearances of the targets are similar, with no distinct inter-specimen heterogeneity. They each exhibit at least one long fracture that is greater than half the length of the target, a feature that is also commonly seen in XCT of Bennu samples [Ryan et al. in press]. We show example XCT cross sections of different Bennu sample morphologies in Fig. A1. Larger pre-existing fractures lead to weaker targets; therefore, understanding the flaw-size distribution of targets, even in a qualitative sense, is necessary for controlling experimental conditions. We used the micro-XCT data to generate shape models of each target, which provide a precise volume measurement. With measurements of their masses, we computed bulk densities for the targets that range from 1.36–1.41 g/cm$^3$. Using the ideal gas pycnometer at the University of Arizona's Lunar and Planetary Laboratory, we find that a ~1-cm sample of the simulant has a grain density of 2.45 g/cm$^3$. This implies that the targets have microporosities, $\phi$ = 42.4–44.5%, which falls within the range for centimeter-scale Bennu samples ($\phi$ = 39–45%) [Ryan et al. in press]. We performed the experiments at the HyFIRE facility at the Johns Hopkins University's Hopkins Extreme Materials Institute (Fig. 7a,b). The experimental setup is described in the Appendix Sec. A, and the experimental conditions are summarized in Table 1.

| Experiment Designation | $M_T$ (g) | $V_t$ (cm³) | $\rho$ (g/cm³) | $\phi$ (%) | $M_P$ (g) | $\delta$ (g/cm³) | $U$ (km/s) | $M_{LF}/M_T$ |
|---|---|---|---|---|---|---|---|---|
| **CI_LG_HyFire_001** | 966 | 694.2 | 1.39 | 43.1 | 0.039 | 2.79 | 5.31 | 0.554 |
| **CI_LG_HyFire_002** | 1084 | 792.1 | 1.37 | 44.0 | 0.039 | 2.79 | 5.33 | 0.539 |
| **CI_LG_HyFire_003** | 962 | 709.1 | 1.36 | 44.5 | 0.003 | 7.90 | 5.39 | 0.996 |
| **CI_LG_HyFire_004** | 1002 | 712.1 | 1.41 | 42.4 | 0.177 | 2.79 | 5.34 | 0.044 |

**Table 1.** Summary of impact experiments into simulant targets. $M_t$: target mass (0.01% uncertainty), $V_t$: target volume (1.5% uncertainty), $\rho$: target bulk density (1.5% uncertainty), $\phi$: target microporosity (1.5% uncertainty), $M_p$: projectile mass (3% uncertainty), $\delta$: projectile density (3% uncertainty), $U$: impact speed (0.1 m/s uncertainty), $M_{LF}$: mass of the largest fragment (0.3% uncertainty).

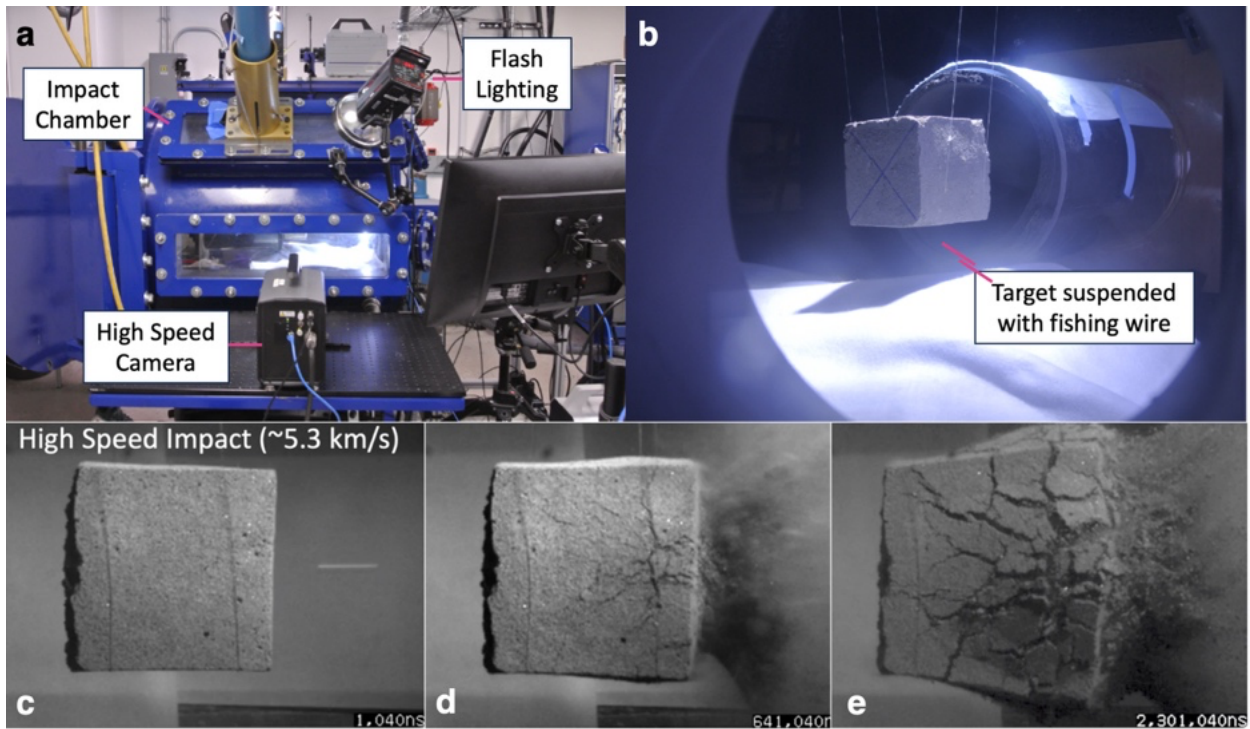

**Figure 7. a**, Impact chamber and diagnostics setup at the Hopkins Extreme Materials Institute's HyFIRE facility. **b,** A target suspended with fishing wire. **c–e,** Example images from experiment CI_LG_HyFire_004, one of three head-on impact experiments at ~5.3 km/s. **c,** The target right before impact (0.001 ms after firing). The projectile is the white streak on the right. **d,** ~0.64 ms after impact, the target begins to fracture. **e,** ~2.3 ms after impact, the target is fully disrupted. The largest post-impact fragment from this impact is 4.4% the mass of the target.

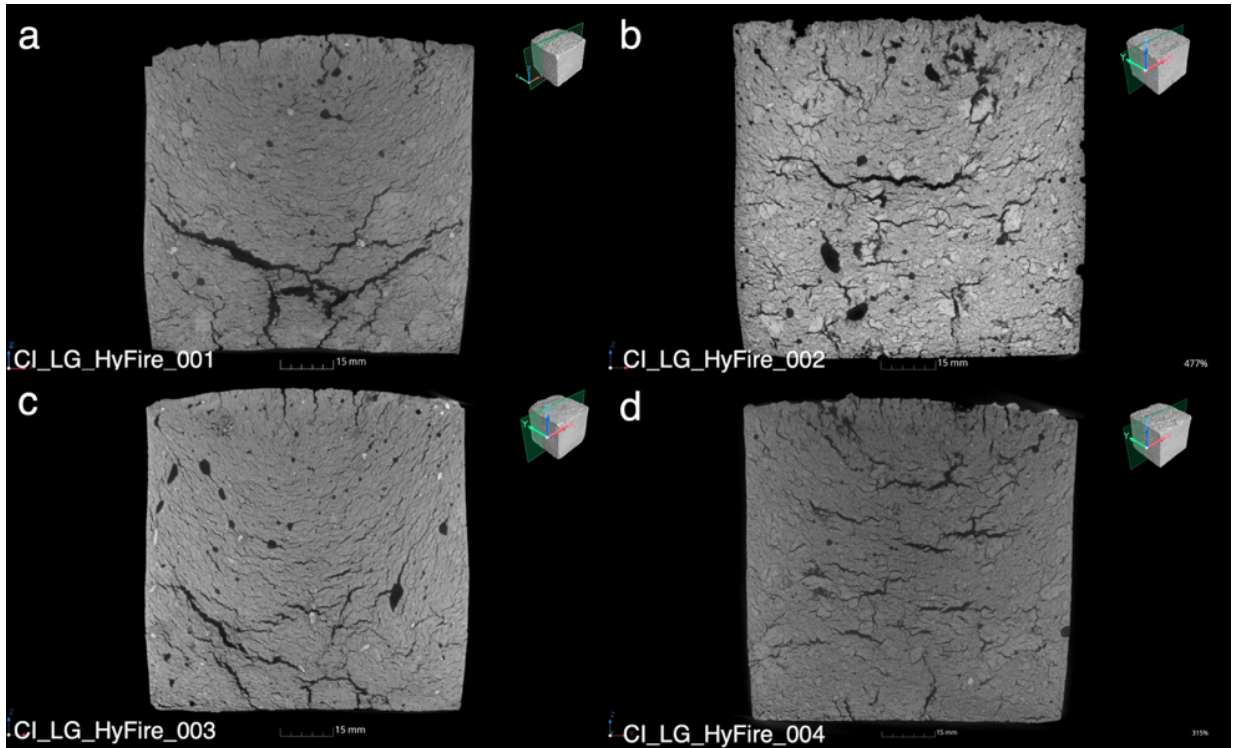

**Figure 8.** Pre-impact micro-XCT slices of simulant targets at voxel sizes of ~50 μm, which show qualitatively similar interior structures. The insets show the location of each slice in the target. **a,** CI_LG_HyFire_001, which has a microporosity, $\phi$ = 43.1%. **b,** CI_LG_HyFire_002 with $\phi$ = 44.0%. **c,** CI_LG_HyFire_003 with $\phi$ = 44.5%. **d,** CI_LG_HyFire_003 with $\phi$ = 42.4%.

*3.2 Experimental Outcomes: Mass Loss, Ejecta Velocity, and Post-Impact XCT of Craters*
In total, we performed four impact experiments that span the range of impact outcomes, from cratering (experiment CI_LG_HyFire_003), to catastrophic disruption (where the largest fragment has a mass, $M_{LF}$, that is ~50% of the mass of the target, $M_T$; experiments CI_LG_HyFire_001 and CI_LG_HyFire_002), to super-catastrophic disruption (where $M_{LF} \ll 50\%$; experiment CI_LG_HyFire_004), and measured the fragment mass distribution for each of these. Fig. 7c-e shows an example impact sequence (experiment CI_LG_HyFire_004), showing the gradual disruption of the target. For the disruptive experiments, we measured the masses of the ~100 largest fragments that were retrieved from the impact chamber (Fig. 9a). Although the material is relatively weak, we verified through low-frame-rate videography that the fragments were formed from the impact by the projectile and not any subsequent impacts with the chamber. The fragments are generally rough in texture (Fig. 9b), morphologically resembling the hummocky Bennu samples; some fragments have smooth and angular faces, but these are likely an artifact of the simulant preparation process (sourced from the outer edge of the smoothed cube targets).

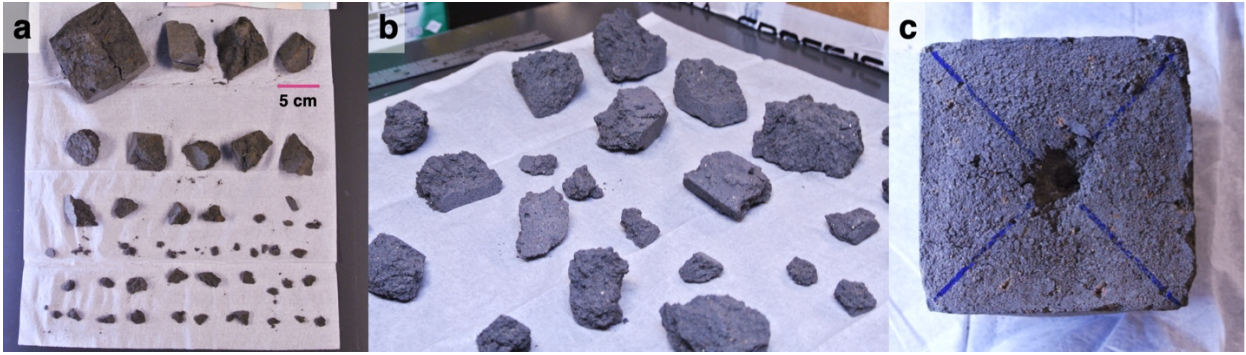

**Figure 9.** Hand-held camera images of the products of the hypervelocity impact experiments. **a,** Post-impact fragments of a disruptive impact (CI_LG_HyFIRE_001), where the largest fragment has a mass of ~ 50% of the original target. **b,** Oblique view of impact fragments that highlights their morphologies from the super-catastrophic disruption impact (CI_LG_HyFIRE_003). **c,** Top-view of cratering impact (CI_LG_HyFIRE_004) that shows crater morphology.

Fig. 10 shows the cumulative mass frequency distribution (CMFD) of fragments from the three disruptive experiments. Two of the three experiments were at the catastrophic disruption limit of the material ($M_{LF}/M_T \cong 0.5$), and the third experiment was super-catastrophic with $M_{LF}/M_T < 0.05$ (Table 1). We modeled the CMFD of these experiments as a power-law distribution, where the cumulative number, $N$, of fragments with mass $M_F$ normalized by $M_T$, is given by

$$N\left(> \frac{M_F}{M_T}\right) = b \left(\frac{M_F}{M_T}\right)^{-a} \quad (2)$$

We performed a least-squares regression fit to the CMFD data with Eq. (2) and find that the disruptive experiments have similar values of $a = 0.69 \pm 0.07$ and $0.63 \pm 0.10$, whereas the super-catastrophic disruption experiment has a steeper slope with $a = 1.42 \pm 0.19$. The CMFD power-law exponent, $a$, can be converted to a cumulative size frequency distribution (CSFD) power-law exponent, $s$, for clearer comparison with those of asteroid boulders and samples through the following relation: $s = 3a$. As such, the fragments from the catastrophic- and super-catastrohic disruption cases would have equivalent values of $s = 2.07 \pm 0.21$ and $1.89 \pm 0.30$, shallower than that found for Bennu globally ($s \sim 2.9$) [DellaGiustina & Emery et al. 2019] but may more closely match that of the OSIRIS-REx sampling site and the sample itself ($s \sim 2.3$) [Burke et al 2021, Lauretta & Connolly et al. 2024, Ryan et al. in press]. The super-catastrophic disruption case has $s = 3.69$–$4.83$, which more closely matches that of Bennu and other NEAs [e.g., DellaGiustina & Emery et al. 2019, Pajola et al. 2024]. These results support that collisional processes are a plausible driver of the boulder size distribution on NEAs. However, it is difficult to conclude this firmly without more experiments to establish the reproducibility of the results. Nevertheless, we note that a similar transition from shallower slopes of $a \sim 0.5$ for less disruptive impacts to steeper slopes of $a \sim 1.5$ for more disruptive impacts was observed in impact experiments on basalt targets [Fujiwara et al. 1977, Takagi et al. 1984].

In our experiments, we varied the range of specific impact energies, $Q$, which is the impact kinetic energy normalized by the total mass of the impactor and target [Fujiwara et al. 1977]. As noted previously, the catastrophic disruption threshold, $Q^*_S$, is defined as the value of $Q$ which results in the largest fragment having half of the total original mass of the system. Typical values of $Q^*_S$ for rocks, ices, and metals can be found in Fujiwara et al. [1977] and Holsapple et al. [2002]. Our experiments probe $Q$ = 45 to 2522.7 J/kg, with the two near-disruptive cases having $Q$ = 512 and 570 J/kg. Fitting a simple linear relationship between $M_{LR}/M_T$ and $Q$ for the disruptive impacts yields $Q^*_S$ = 724 J/kg. This value is less than half that reported for some ordinary chondrite meteorites and the Allende CV chondrite [e.g., Flynn et al. 2018, Flynn et al. 2025], but close to that obtained in experiments on carbonaceous chondrites, namely Aguas Zarcas (CM2) and NWA 4502 (CV3), which have porosities of ~20% and ~2.1%, respectively. It is not clear how the interiors of those targets or impact conditions (e.g., impact angle) may have influenced the outcomes. This result highlights the overall weak nature of Bennu analogs compared to other asteroid materials, which was hypothesized from remote sensing analysis of Bennu boulders [e.g., Ballouz et al. 2020, Rozitis et al. 2020].

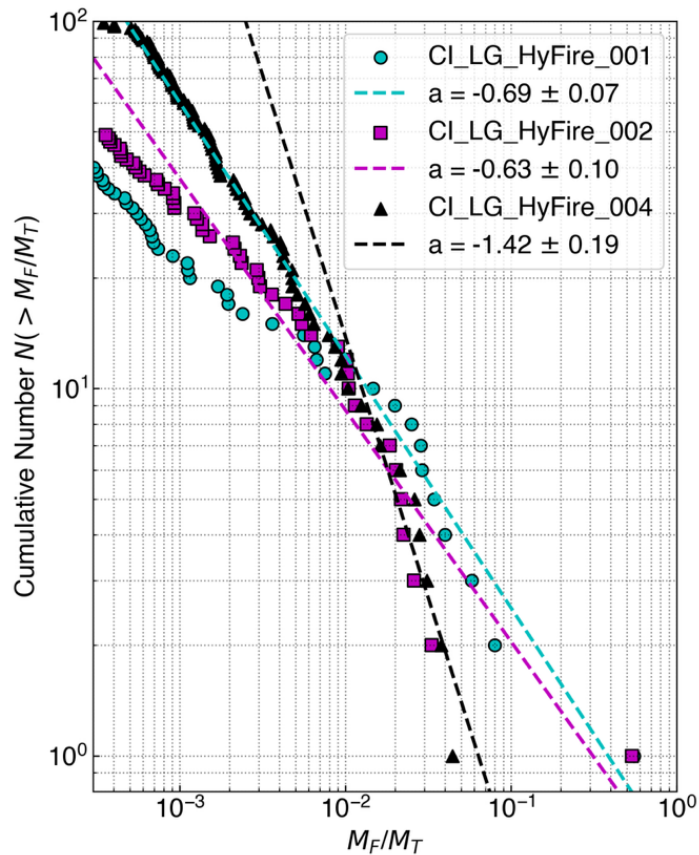

**Figure 10**. The cumulative mass frequency distribution (CMFD) of fragments produced by the three disruptive impact experiments. The data points show the cumulative number of fragments with normalized mass $\geq M_F/M_T$. The dashed lines show the best-fit power-law curve (Eq. 2) for the corresponding data points of the same color. The legend gives the best-fit power-law exponent $a$ for each curve.

We used the top-view videography for the disruptive experiments CI_LG_HyFire001 and CI_LG_HyFire002 to estimate the post-impact velocity of the largest fragments in each experiment. Fig. 11 shows the first and last frame used for this analysis for CI_LG_HyFire001. For each frame, we used Meta's Segment Anything [Kirillov et al. 2023] to identify the largest fragment and isolate its edge. Then we took the median value of the edge's vertical position and fit a linear relationship to its time evolution. Using our calibration images to convert pixel scales to physical units, we found that the largest fragment produced in CI_LG_HyFire001 moves at a constant velocity of 0.46 ± 0.03 m/s over 300 frames (which is equivalent to a total elapsed time of 15 ms). The error is estimated from a 2-pixel position uncertainty, which dominates over the linear fit to the block's frame-to-frame motion (<0.5%). For CI_LG_HyFire002, the largest fragment moves at a constant velocity of 0.73 ± 0.03 m/s.

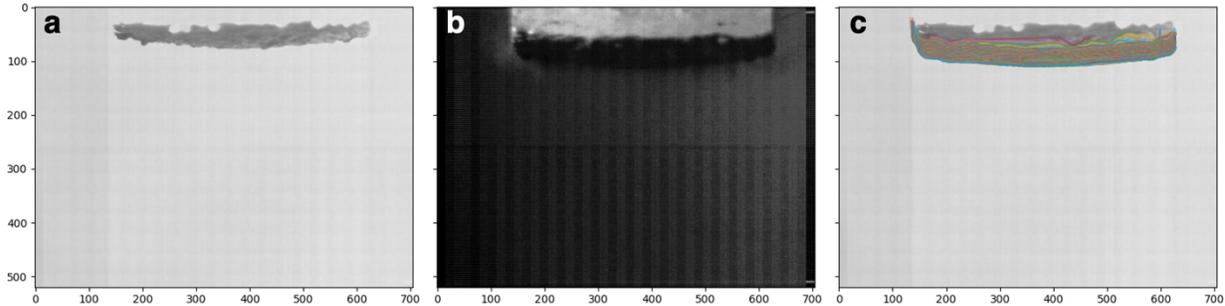

**Figure 11.** Sequence of frames for experiment CI_LG_HyFire001, looking down into the chamber (the impact direction is from the top of the page toward the bottom). The images show the largest fragment of this disruptive impact. **a,** First frame in sequence that is over-exposed due to the flash used to capture the impact for the higher-frame-rate side-view cameras. **b,** Final frame, 15 ms after **a**. **c,** Edges of the fragment in each frame, shown here super-imposed on the first frame.

The extremely low post-impact speed of the largest fragments compared to the speed the disruptive projectile (~four orders of magnitude difference) suggests that the dynamics of impact-generated fragments need to be evaluated carefully to better understand their fate, even on small micro-gravity bodies. For example, Bennu's escape velocity is only ~0.2 m/s, and the asteroid's surface has been shown to be dissipative, with a directly measured coefficient of restitution of 0.57 ± 0.01 [Chesley et al. 2020]. For the scenario of a fragment impacting a rigid surface at a speed of 0.46 m/s and 0.73 m/s, those fragments would rebound with speeds of 0.26 m/s and 0.42 m/s, respectively, which would lead to their escape from Bennu. As we discuss in Sec. 5, these fragments may penetrate into a porous regolith, rather than rebound off the surface.

Figs. 9c and 12 show the outcome of non-disruptive experiment, CI_LG_HyFire_003, which resulted in an impact crater with the largest remnant (that is, the target) retaining >99% of the original mass. To characterize the crater and assess whether any other physical modifications occurred, we conducted a post-impact micro-XCT scan (Fig. 12). The crater has a similar morphology to craters on Bennu samples described in Section 2, with an outer spall diameter of 25 ± 0.1 mm and an inner pit diameter of 10 ± 0.1 mm. The ratio of spall-to-pit diameter, 2.5 ± 0.03, falls within the range observed for terrestrial porous brittle materials (~1.5–3.0) such as gypsum, tuff, and sintered glass beads [Nakamura 2017]. Delbo et al. [2022] reported a 3.23-m-diameter crater on a Bennu boulder that has a similar spall morphology. The crater on that boulder has a $d/D = 0.44$ and a spall-to-pit diameter of 2.2 — values similar to that seen in the sample and the simulant. Delbo et al. [2022] interpreted the morphological properties of the crater to signify a density or strength stratification of the boulder, due to its similarity to bench craters seen on the surfaces of larger bodies and in laboratory impact experiments [Hörz & Cintala 1997]. Our observations suggest that these spallation features are more likely a natural consequence of the bulk mineralogical properties of Bennu material.

We measured the crater volume, $V_c = 1.44 \pm 0.03$ cm$^3$, and observed that the material in the vicinity of the impact region had been compacted. The region of pore compaction extends to 12–13 mm from the point of impact and surrounds the region of the impact pit. The extent of this compacted region is equivalent to the radius of the spall. In comparison, the higher XCT attenuation region near the floor of the crater on OREX-800047-0 shown in Fig. 5b extends to ~90% of that crater's spall radius. We find no other changes to the target's pore structure. This observation suggests that impacts into Bennu material can lead to compaction cratering, which would suppress the formation of some ejecta. This finding contradicts previous experimental work suggesting that compaction cratering should only occur on porous (≳40%) planetary bodies with diameters ≳ 30 km [Housen et al. 2018]. The discrepancy between our observation and that work is likely due to the different target materials; Housen et al. [2018] used a sand-perlite and dry granular pumice). Our work is consistent with the results of impact experiments on porous sandstone targets (23% porosity), which showed localized pore-space compaction near the crater floor [Buhl et al. 2013]. However, our finding supports the hypothesis of Cambioni et al. [2021] that the production of fines on Bennu is suppressed due to compaction cratering. For the phyllosilicate-rich materials that compose the bulk of the Bennu samples [Lauretta & Connoly et al. 2024], the submicron-scale porosity [Ryan et al. in press] likely plays a significant role in enabling compaction cratering.

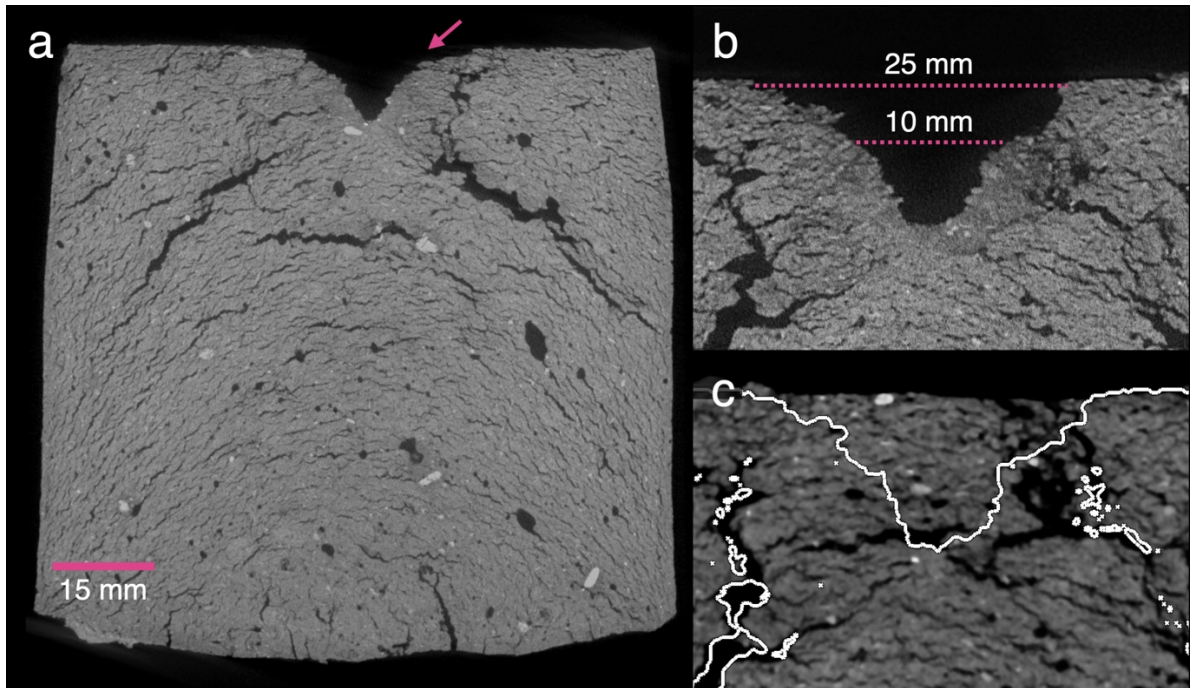

**Figure 12. a,** Micro-XCT slice of the post-impact target from experiment CI_Lg_HyFire_003 at 51 μm/voxel. The crater is highlighted by the magenta arrow. **b,** Close-up showing the morphology of the crater. The dotted lines show the outer and pit diameter of the crater. **c,** Outlines of the void edges (white) from the post-impact XCT scan, overlain on the pre-impact XCT scan to highlight the compaction of the porous structure from the impact.

The laboratory impact experiments described in this section allowed us to calibrate material parameters for a smooth-particle-hydrodynamics (SPH) code that can scale up experimental results to evaluate impact outcomes for boulder-sized targets (tens of meters).

# 4. Simulations of Hypervelocity Impacts into Bennu Boulders

To gain a more complete understanding of the post-impact kinematics of impact fragments, we used the experimental results described in the previous section to calibrate numerical models to scale our results to boulder-sized targets (tens of meters). We used the SPH code *miluphCUDA* [Schäfer et al. 2016], which is capable of using different equations of state (EOS) as well as material strength and damage models. Details of the simulation setup can be found in Appendix Section A2, and Table 2 summarizes simulation parameters. The scalar damage parameter, *d*, characterizes the influence of cracks in a given volume and ranges from 0 (undamaged, intact material) to 1 (fully damaged material that cannot undergo any tension or deviatoric stress). A critical aspect of the simulations is that a pre-impact target is initialized, in part, by using a Weibull distribution [Weibull 1939] to describe the number of flaws per unit volume, *n*, with failure strains that are lower than a strain, *ε*, such that:

$$n(\epsilon) = k\epsilon^m, \quad (3)$$

where *k* and *m* are the two Weibull parameters. We developed Weibull parameter values using the results of our impact experiments.

| Property | Value | Reference |
|---|---|---|
| Shear Modulus (Pa) | 3.74e9 | Hildebrand et al. (pers. comm., 2025) |
| Grain density, $\rho_g$ (g/cm3) | 2.45, 2.8 | This work, Ryan et al. (in press) |
| Tillotson: A (Bulk Modulus, Pa) | 5.04e9 | Hildebrand et al. (pers. comm., 2025) |
| Tillotson: B (Pa) | 0.053e9 | Nakamura et al. (2022) – Ryugu "soft" |
| Tillotson: $E_0$ (J/kg) | 0.2e6 | Nakamura et al. (2022) – Ryugu "soft" |
| Tillotson: $E_{iv}$ (J/kg) | 4.5e6 | Nakamura et al. (2022) – Ryugu "soft" |
| Tillotson: $E_{cv}$ (J/kg) | 14.5e6 | Nakamura et al. (2022) – Ryugu "soft" |
| Tillotson: a | 0.5 | Nakamura et al. (2022) – Ryugu "soft" |
| Tillotson: b | -0.457 | Nakamura et al. (2022) – Ryugu "soft" |
| Tillotson: $\alpha$ | 5.0 | Nakamura et al. (2022) – Ryugu "soft" |
| Tillotson: $\beta$ | 5.0 | Nakamura et al. (2022) – Ryugu "soft" |
| p-alpha: $p_e$ (Pa) | 1.0e6 | Jutzi et al. (2009) – pumice |
| p-alpha: $p_t$ (Pa) | 6.80e7 | Jutzi et al. (2009) – pumice |
| p-alpha: $p_c$ (Pa) | 2.13e8 | Jutzi et al. (2009) – pumice |
| p-alpha: $\alpha_0$ | 1.75 | Jutzi et al. (2009) – pumice |
| p-alpha: $\alpha_e$ | 4.64 | Jutzi et al. (2009) – pumice |
| p-alpha: $\alpha_t$ | 1.90 | Jutzi et al. (2009) – pumice |
| p-alpha: $n_1$ | 12.0 | Jutzi et al. (2009) – pumice |
| p-alpha: $n_2$ | 3.0 | Jutzi et al. (2009) – pumice |

**Table 2.** Summary of the Tillotson EOS and p-alpha model parameters used in our simulations and each value's associated reference.

*4.1 Calibration Simulations for Bennu Material*

We undertook a suite of simulations to calibrate the SPH code for impacts into Bennu-like material, with the goal of replicating the results of the disruption experiments. This set of simulations consisted of $N \sim 730,000$ SPH particles for the target, with a smoothing length of 2 mm, that make up a cubic target with each side having a length of 9 cm. The total simulated time, $t$, is 5 ms, which is sufficient to fully capture the fragmentation process as the number of sound crossings, $N_v > 100$, with $N_v = c_s t / D_{targ}$. $c_s$ is the material sound speed, and $D_{targ}$ is the target diameter.

As suggested by Asphaug et al. [2002] and discussed further in Jutzi et al. [2009], there is some degeneracy in impact outcomes for different combinations of Weibull parameters, $m$ and $k$, as fragmentations are controlled by the strength scaling parameter $\Phi = \ln(kV_T) / m$, where $V_t$ is the volume of the target. As such, an impact into a target with constant $\Phi$ should result in a self-similar outcome. Following Jutzi et al. [2009], we set $m = 9.5$ and varied the value of $k$. We explored $k$ values from $10^{40}$ to $10^{42}$ cm$^3$ to find impact outcomes that closely matched our experimental results. Table 3 summarizes the *miluphCUDA* simulations that were conducted.

Fig. 13a,b shows a side-by-side comparison of impact experiment CI_LG_HyFire_001 (Table 1), which led to the catastrophic disruption of the 9-cm simulant cube, and a *miluphCUDA* simulation, both at 2.5 ms after impact. The simulation in Fig. 13b used the parameters in Table 2 with $m = 9.5$ and $k = 10^{40}$ cm$^3$. The comparison shows that the SPH simulation can reproduce broadly similar features to the experiments.

| Sim # | Material | $R_T$ (cm) | $\rho$ (g/cm$^3$) | $\phi$ (%) | $M_P$ (g) | N | m | k |
|---|---|---|---|---|---|---|---|---|
| 1 | EC | 4.5 | 1.4 | 43 | 0.039 | 7.30e5 | 9.5 | 8e37 |
| 2 | EC | 4.5 | 1.4 | 43 | 0.039 | 7.30e5 | 9.5 | 1e40 |
| 3 | EC | 4.5 | 1.4 | 43 | 0.039 | 7.30e5 | 9.5 | 1e41 |
| 4 | EC | 4.5 | 1.4 | 43 | 0.039 | 7.30e5 | 9.5 | 5e41 |
| 5 | EC | 4.5 | 1.4 | 43 | 0.039 | 7.30e5 | 9.5 | 1e42 |
| 6 | Bennu | 2500 | 1.4 | 43 | 2e6 | 3.03e5 | 9.5 | 5e41 |
| 7 | Basalt | 2500 | 2.7 | - | 2e6 | 3.03e5 | 8.5 | 5e34 |

**Table 3.** Summary of SPH simulations. The first five simulations were for the experiment-based calibration (EC) effort described in Sec. 4.1 that attempted to replicate the outcomes of the disruptive experiments described in Sec. 3. All simulations were set up to for the projectile to impact the target head-on at a speed of 5.3 km/s. $R_t$: target radius, $\rho$: bulk density, $\phi$: target microporosity, $M_p$: projectile mass, $N$: number of particles in target, $m$ and $k$: Weibull parameters (see Eq. 3).

To assess which values of Weibull parameters best reproduce the experiments, we compared the simulation fragment mass frequency distribution to that of the experiments (as shown in Fig. 13c). In the simulations, the SPH particles of individual fragments are not well separated as to allow for segmentation through typical clustering algorithms. Therefore, we employed a friends-of-friends algorithm [e.g., Benz & Asphaug 1994] to identify individual fragments as clusters of SPH particles separated from the rest by a layer of damaged particles. The mass frequency distribution of the SPH simulations could then be constructed and compared to the experimental results, which we show in Fig. 13c. For this comparison we computed the average normalized mass of the fragments for the two disruptive experiments (CI_LG_HyFire_001 and CI_LG_HyFire_002). We found that the best match to the experimental results to be the simulation with Weibull parameters $m = 9.5$ and $k = 5 \times 10^{41}$ cm$^3$.

We used our values for the Weibull parameters to estimate the equivalent tensile strength of the material ($\sigma_t$) via:

$$\sigma_t \approx (kV_T)^{-\frac{1}{m}} E_S, \qquad (4)$$

where $E_S$ is the target's Young's modulus. Using measured shear and bulk moduli values, we estimated $E_S \sim 9$ GPa. Thus, $\sigma_t \sim 0.18$ MPa for 9-cm cubes, and the compressive strength, $\sigma_c = 1.5$–$2.8$ MPa, based on empirical relationships observed in rocks for the relationship between $\sigma_t$ and $\sigma_c$ [Zhang 2016]. These strength values are commensurate with the generally weak nature of Bennu material [e.g., Ryan et al. in press, Lauretta & Connolly et al. 2024]. Using the same Weibull parameters, we also estimated $\sigma_t = 41$ kPa and $\sigma_c = 0.33$–$0.62$ MPa for a 1-m-diameter Bennu boulder. The range in $\sigma_c$ is within that estimated for Bennu's boulders based on remote characterization of craters (0.44–1.7 MPa) [Ballouz et al. 2020].

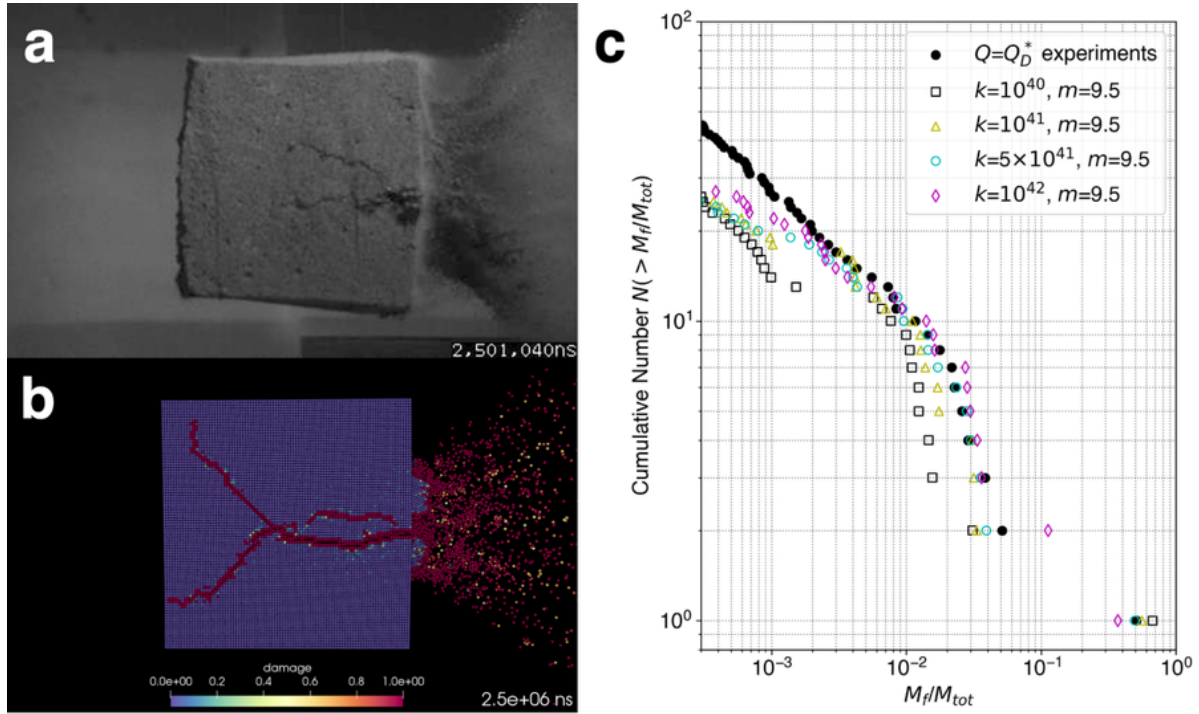

**Figure 13.** Comparison of a disruptive impact experiment (CI_LG_HyFIRE_001) (**a**) to a SPH simulation (Sim #4) (**b**) at 2.5 ms after impact. The extent of damage (scalar damage parameter, *d*) in the simulation is indicated by color, with red indicating fully damaged. **c**, We conducted impact simulations that matched laboratory experimental conditions, varying the Weibull parameter, *k* (e.g., [Benz & Asphaug 1994]), and compared the CMFD results of simulations to the experiments, finding a best correspondence for $m = 9.5$ and $k = 5 \times 10^{41}$ cm$^3$. The number of SPH particles in each simulation is ~ 730,000.

*4.2 Simulations of Bennu Boulder Disruption*

We ran SPH simulations (parameters summarized in Table 3) of impacts into a 50-m-diameter Bennu boulder to evaluate the fragmentation dynamics. We chose the 50-m target diameter because it is on the larger end of the boulder size range on Bennu [DellaGiustina & Emery et al. 2019], and we are interested in how these large boulders may be sources of finer regolith. We also ran a simulation for an impact into a 50-m-diameter basaltic boulder as a point of comparison, as the response of basalt is well understood [e.g., Benz & Asphaug 1994, Nakamura & Fujiwara 1991]. We used the equation of state and Weibull parameters for basalt from Benz & Asphaug [1994]; unlike the Bennu boulder target, we did not use a pore-compaction (p-alpha) model [Jutzi et al. 2009]. The impact velocity was 5.3 km/s, the average impact speed in the main asteroid belt [e.g., Bottke et al. 2005]. Each simulation was run for 0.2 s, which corresponds to $N_v > 10$. We used the friends-of-friends algorithm to find all the individual fragments in the simulation, enclosed by fully damaged particles, as a function of time. Figure 14 shows the evolution of $M_{LF}/M_T$ in a non-dimensional form ($N_v$).

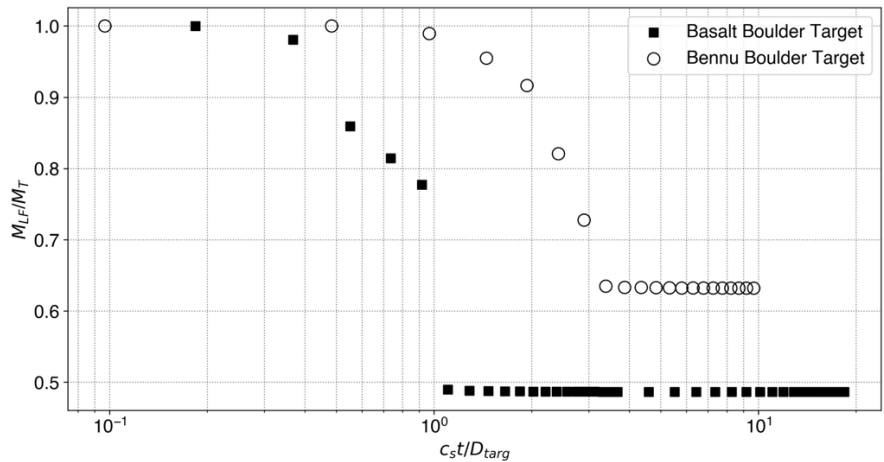

**Figure 14.** The simulations of disruptive impacts into 50-m-diameter targets of different material types at 5.3 km/s were run for sufficiently long (0.2 s) that the mass of the largest remnant ($M_{LF}/M_T$) converges. The data points show the evolution of $M_{LF}/M_T$ as a function of a *p*-wave crossing time, $c_s t/D_{targ}$, where $c_s$ is the material sound speed, $t$ is simulation time, and $D_{targ}$ is the target diameter.

Fig. 15a shows a sequence of snapshots from the Bennu boulder (top) and the basalt boulder (bottom) simulations. The different physical properties of the targets lead to unique fragmentation patterns across their interior. The intact core and the spallation of the antipode of the basalt target is a feature that is commonly seen in both experimental and numerical work [e.g., Nakamura & Fujiwara 1991, Benz & Asphaug 1994]. Contrastingly, the Bennu boulder's core does not remain intact. It responds to the impact by forming a deep transient crater and linear fractures. Despite the extensive damage to its interior, it retains a larger fraction of its mass compared to the basalt (see Fig. 14). This difference is likely due to exclusion of pore compaction in the latter. As shown by, e.g., Jutzi et al. [2009], inclusion of the effects of pore space can lead to porous objects being more resistant to disruption in simulations. This highlights the importance of appropriately including pore space effects in simulations. We neglect it here for the basalt case to demonstrate the reproducibility of experimental outcomes and prior computational work [Nakamura & Fujiwara 1991, Benz & Asphaug 1994].

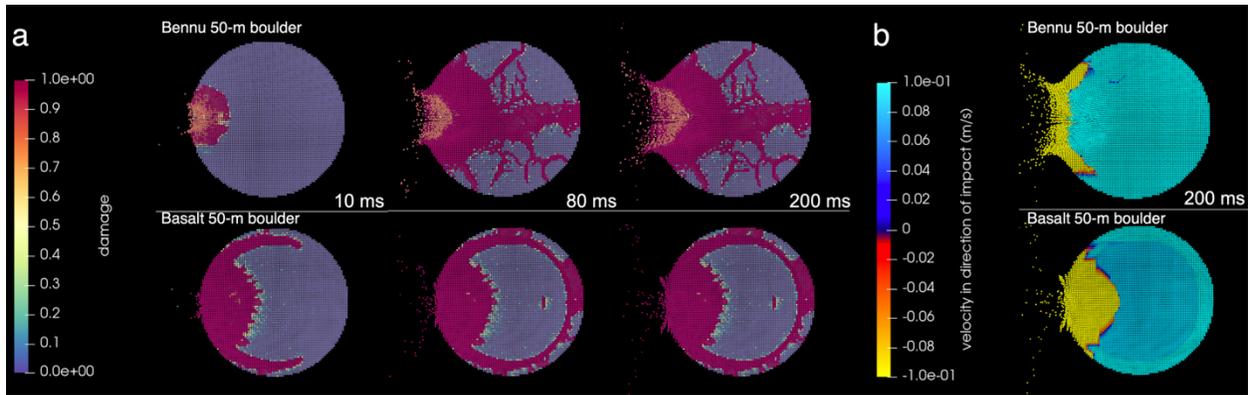

**Figure 15. a**, Results of SPH simulation of disruptive impacts at 5.3 km/s into 50-m-diameter Bennu (top) and basalt (bottom) boulders, demonstrating the propagation of damage in the interior of the targets with time. Each sequence shows a slice through the middle of the target co-planar with the impactor's velocity, with the impactor having traveled from the left to the right. The SPH particles are colored by their damage, with red indicating fully damaged (damage parameter = 1). **b**, The final snapshot of each simulation (at 200 ms) with the SPH particles colored by their speed along the direction of impact (bluer colors, same direction as projectile; warmer colors, opposite direction of projectile).

For both boulders, despite the differences in their physical properties, the majority of their mass moves in the same direction as the projectile (Fig. 15b, bluer colors). A small fraction of the disrupted targets is ejected in the opposite direction (Fig. 15b, warmer colors). The magnitude of the ejection velocity is small relative to the impact speed (5.3 km/s), with all the mass moving at speeds less than a few tens of centimeters per second, similar to what was observed in our laboratory impact experiments.

Fig. 16 shows rose diagrams for the post-impact target's normalized mass (mass traveling in a direction divided by the total target mass), speed, and normalized momentum (momentum divided by the impactor's momentum). For both targets, as also shown in Fig. 15b, the majority of the target's mass is directed along the impact direction (Fig. 16a, d), despite the fact that the fragments with the highest speeds (Fig. 16b, e), which carry a significant amount of momentum (Fig. 16c, f), are in the opposite direction. The fast-moving ejecta provide additional momentum to the largest fragments through a jetting effect, which can be described as a momentum enhancement factor [e.g., Stickle et al. 2017]. The precise value of the momentum enhancement factor is the subject of numerous experimental and numerical impact experiments, as it is important for the formulation of planetary defense mitigation strategies [e.g., Stickle et al. 2017].

A potentially contributing factor to momentum enhancement for carbonaceous chondrites is impact-induced devolatilization of shocked material [Kurosawa et al. 2025], which we do not include in our numerical models. Kurosawa et al. [2025] suggested that shocked carbonaceous chondrites may preferentially escape their host asteroids, which may explain the dichotomy in shock degree between carbonaceous chondrites and ordinary chondrites. From our limited set of experiments, it appears that the overall kinematics of fragmented material is well captured in our numerical simulations; however, further work on impact-induced devolatilization of hydrated asteroids may shed light on potential divergent surface evolution pathways between carbonaceous and stony asteroids.

For impacts on a boulder on an asteroid, these results signify that a very large fraction of the mass is driven into the asteroid's surface. In the Bennu boulder simulation, 54%, 72%, and 90% of the total normalized mass is directed within 30°, 45°, and 80° of the impact direction, respectively. In the basalt boulder simulation, 72%, 79%, and 85% of the total normalized mass is directed within 30°, 45°, and 80° of the impact direction, respectively. The median fragment speeds are 0.43 m/s for the Bennu boulder and 0.1 m/s for the basalt boulder (Fig. 16b,e). The median fragment speed for the Bennu boulder simulation agrees with that seen in our laboratory experiments, which had a median of 0.6 m/s for the two experiments that were close to the catastrophic disruption threshold.

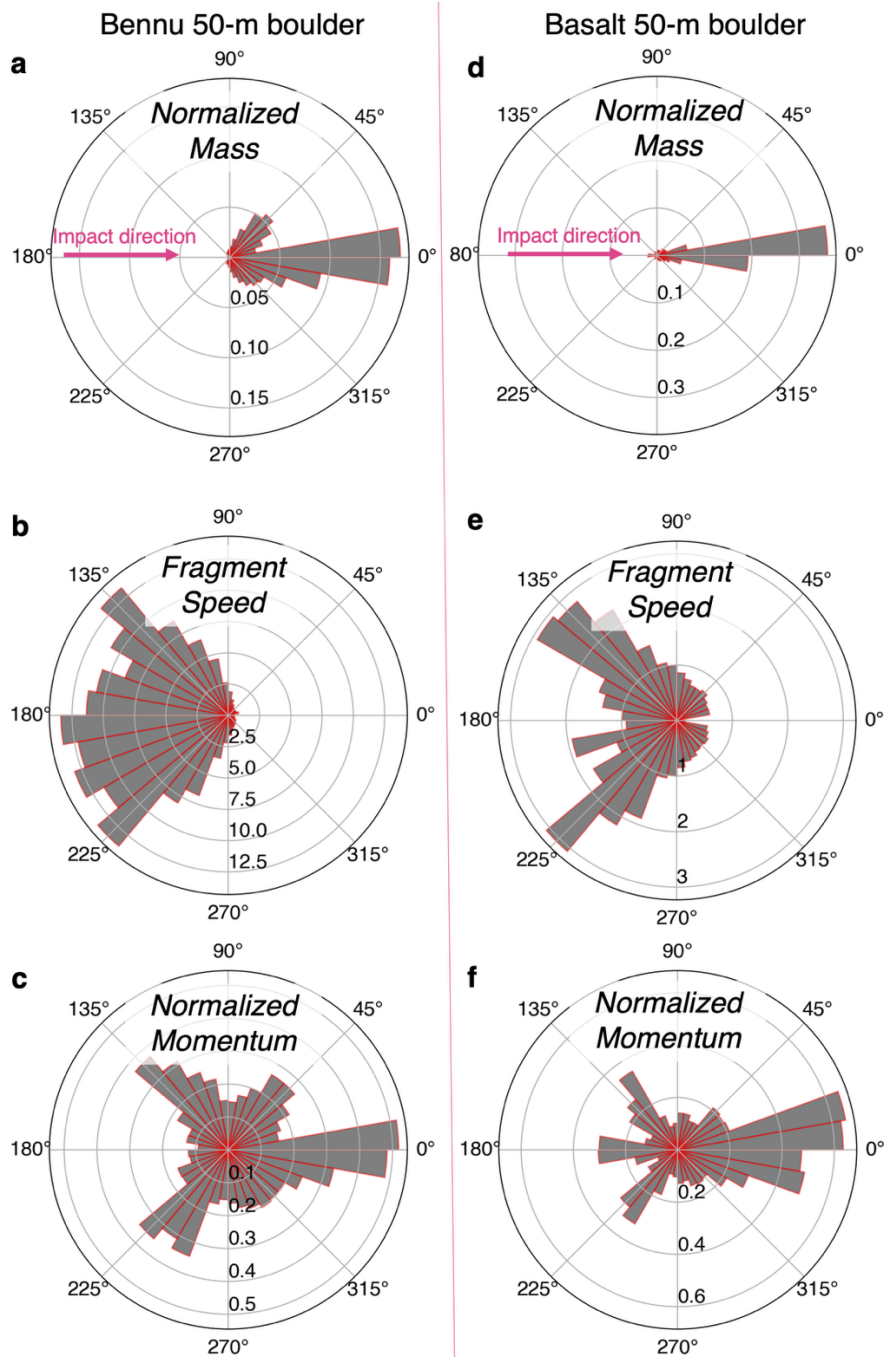

**Figure 16.** Rose diagrams showing the distribution of ejection angles for impact fragments relative to the velocity vector of the impactor (0°) for simulations of impacts disrupting a 50-m-diameter Bennu boulder (left column) and a 50-m-diameter basalt boulder (right column). **a,** The distribution of mass, normalized by the total mass of the target, traveling along each ejection angles (binned at increments of 5°). **b,** Same as **a**, but for the speed of ejecta fragments in meters per second. **c,** Same as **a**, but for the momentum of fragments normalized by the impactor's momentum. **d–f,** Same as **a–c**, but for the basalt target.

Meteoroids rarely impact a planetary surface with a normal incidence angle. Rather, the most likely impact angle is ~45° [e.g., Gilbert 1893, Sheomaker 1962, Melosh 1989]. The simulation results we have presented are also relevant to the outcomes of angled impacts, as long as the impact vector aligns with the center-of-mass of the target boulder. The key difference is that the fragment's velocity with respect to the surface of the asteroid is rotated by an amount equal to the impact angle. Therefore, for an impact onto a boulder at an angle of 45°, fragments would be directed into the surface of the asteroid if they had an ejection angle between −135° and +45°. From our Bennu boulder simulation, a 45° impact would lead to ~85% of the total normalized mass being directed into the surface.

We conclude that the disruption of boulders on the surfaces of asteroids can lead to low ejecta speeds (tens of centimeters per second) relative to the impactor velocity. Previous work [e.g., Housen & Holsapple 2011, Nakamura 2017] had focused on the velocity distribution of material that is directed away from the surface, as these studies were chiefly interested in crater formation rather than disruption dynamics and regolith evolution. Our work is distinct in showing that the disruption of meter-scale boulders drives mass into the surface, resulting in the retention of fragments. The problem transforms from a hypervelocity impact that disrupts a boulder into a low-speed impact of the resultant fragments into the underlying regolith.

Next, we present analysis of friction properties, which are essential for understanding the outcome of low-speed collisions into regolith [e.g., Walsh & Ballouz et al. 2022, Ballouz et al. 2021].

## 5. Angle of Repose Measurement and Penetration Dynamics of Impact-Generated Fragments

Analysis of video-documentation of OSIRIS-REx sample curation activities allowed us to estimate the angle of repose, $\theta$, of the regolith returned from Bennu, which we used to evaluate the penetration dynamics of impact-generated fragments into the surface. Angle of repose measurements provide the means to calculate the angle of internal friction and constrain cohesive properties. The angle of internal friction is a measure of the ability of a particulate assemblage to withstand a shear stress. It is a key parameter in models of rubble-pile deformation [e.g., Zhang et al. 2018], surface mass movement [Ballouz et al. 2019], and surface penetration of the OSIRIS-REx Touch-and-Go Sample Acquisition Mechanism (TAGSAM) [Bierhaus et al. 2018, Ballouz et al. 2021, Walsh & Ballouz et al. 2022]. The internal angle of friction is generally equivalent to the angle of repose, i.e., the maximum slope angle formed by a pile of particles, when cohesive forces can be neglected (larger particle sizes).

*5.1 Measuring the Angle of Repose of Bennu Regolith*
It was critical that the angle of repose measurement occur with the least possible amount of atmospheric exposure, as cohesive properties are altered by the absorption of moisture. We observed the pour-out of Bennu material from the TAGSAM onto a baffle-tray system inside the glovebox. The procedure involved two processors at the NASA Johnson Space Center (JSC) curation facility handling the TAGSAM baseplate using a holder (Fig. 17). The processors then tilted the baseplate until the material began to flow into one of the eight sample collection trays [Lunning et al. 2025], allowing us to passively measure the angle of repose. This procedure is analogous to a bed-tilting experiment, a standard method for determining the angle of repose of granular material [Al-Hashemi & Al-Amoudi 2018]. The angle of repose is estimated by measuring the angle at which the material in the box begins to slide.

We installed two GoPro Hero 8 Angle of Repose Measurement Cameras (ARMCams), placed on a Teflon holder on top of the glovebox. The ARMCams captured the pour-out at 24 frames per second in 4k resolution. The pixel scale of the ARMCam images is <0.2 mm/pixel. Figure 17 shows two examples of tilting in the pour-out procedure that resulted in mass movement off the plate and onto the collecting trays. In total, we captured four sets of movements where we can accurately measure the onset of regolith mobilization; we term these mass movement events (MMEs). The mobilized particles in the video-documentation are millimeters to centimeters in size and rest on top of other particles while a finer layer of dust coats the surface of the baseplate. Therefore, the angle of repose measurements are representative of particle-to-particle interactions.

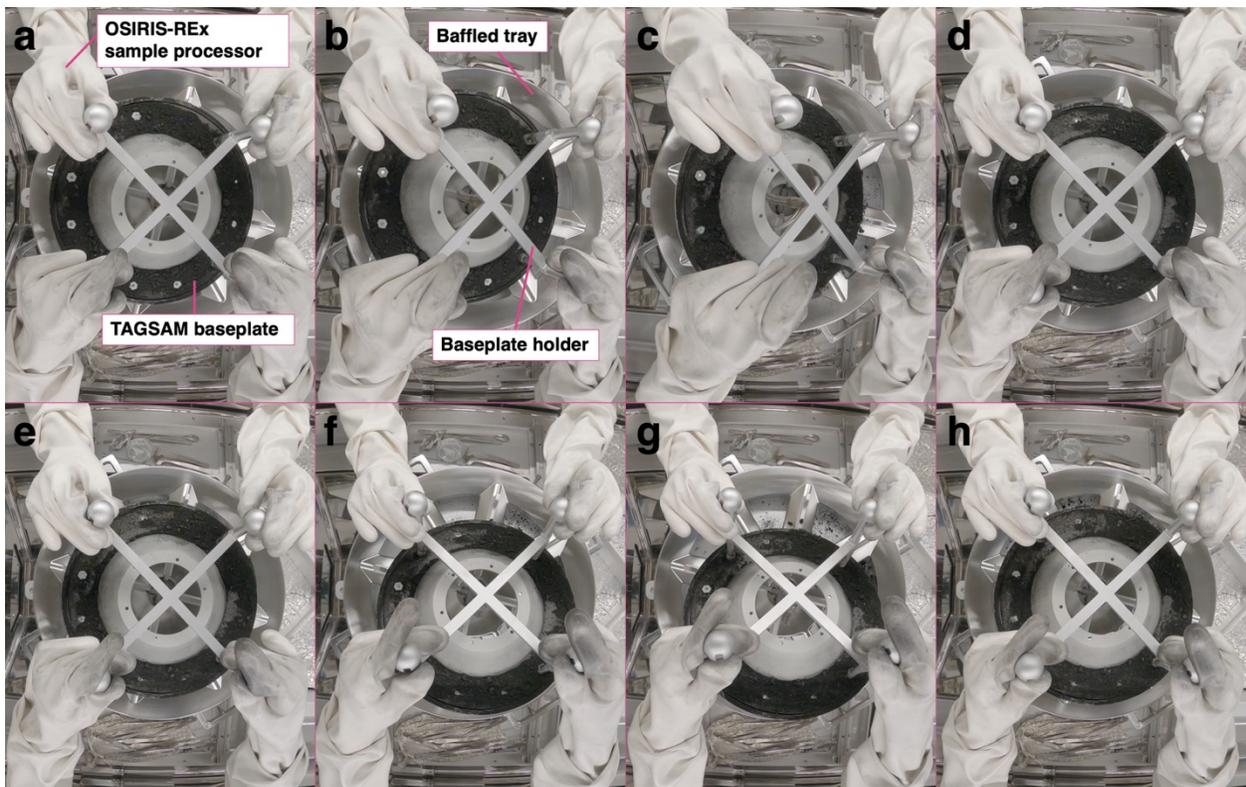

**Figure 17.** Sequence of frames from an ARMCam, capturing the pour-out of Bennu material by NASA JSC processors from the TAGSAM onto a baffled tray system inside the glovebox. The processors installed a four-handled baseplate holder onto the TAGSAM baseplate and then carefully tilted the baseplate until material was mobilized by gravity. **a–d,** The processors tilting the TAGSAM, toward the right in this view, until material begins to move and fall into the baffled tray. **e–h**, As in a–d, but with the tilt direction towards the top of the page.

To obtain an estimate of the angle of the baseplate with respect to gravity (i.e., its slope) at the onset of mass movement, we used a computer vision technique called pose estimation [Bradski 2000]. This method leverages the planar geometry of the baseplate and the baseplate holder. By obtaining the orientation of the backplate at different frames, we can reconstruct the slope at which the regolith begins to slump, which is its angle of repose. For each frame, we identified four key points on the surface of baseplate and the holder, and the pose estimation algorithm then reconstructed the orientation of both objects, which have parallel surfaces. In all cases, the exact instance of surface failure has some uncertainty as the MME is initiated before the tilting motion has come to a stop. Generally, though, the processors would stop a tilt at an angle that is close to that where the MME was initiated. Therefore, the angle of repose is based on the distribution of TAGSAM tilts for a given movement (Fig. 18).

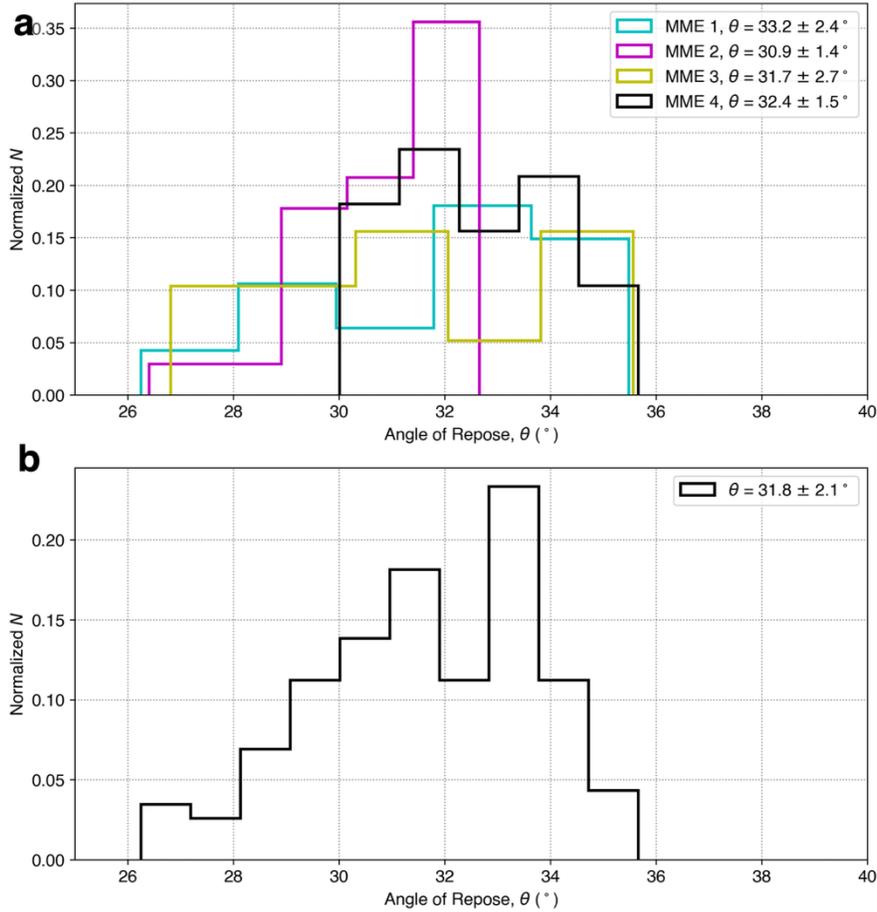

**Figure 18. a,** Distribution of angle of repose, $\theta$, measurements for each mass movement event (MME). **b,** Distribution of angle of repose over all measurements ($N = 123$). We find a median value of $\theta = 31.8° \pm 2.1°$ across the four MME measurements.

Figure 18 shows the that tilt angle at which Bennu material begins to move on the surface of the baseplate ranges from 26° to 36°. The measurements are normally distributed, except for a spike close to 33°, which is likely due to the JSC processors maintaining this tilt that induced sample motion for an extended duration. The median slope for all 123 frames is $31.8° \pm 2.1°$.

If cohesion is negligible for the Bennu material, as suggested by particle-to-particle cohesion measurements [Jardine et al. 2025], then our measurement of the angle of repose is consistent with the angle of friction of 32°–39° estimated based on evidence of mass movement near the OSIRIS-REx sample site [Barnouin et al. 2022]. Our measurement also overlaps within the 1-sigma uncertainty of the angle of internal friction of $31.1° \pm 2.7°$ estimated for Bennu's boulders based on their morphologic characteristics [Robin et al. 2024]. Thus, our pour-out analysis validates inferences from remote sensing.

*5.2 Revisiting Bennu's Near-Surface Bulk Density and Macroporosity*

We revisit penetrometry analysis [Walsh & Ballouz et al. 2022] of the near surface (upper 10–20 cm) of Bennu. Walsh & Ballouz et al. [2022] used spacecraft acceleration data, as it penetrated the surface of Bennu at 10 cm/s for sampling, to determine the bulk density, $\rho_b$; an empirically derived relationship between the peak force from the surface on the spacecraft, $F_p$; the impact speed of the spacecraft, $U_{S/C}$; and $\theta$ yields

$$F_p = 0.62 \tan(\theta)\, \rho_b U_{S/C}^{4/3} \qquad (5)$$

Using values of $F_p$ = 10–15 N and $U_{S/C}$ = 10 cm/s, our newly measured value of $\theta$ gives $\rho_b$ = 580–870 kg/m³, which is in general agreement with that previously determined (440–600 kg/m³ [Walsh & Ballouz et al. 2022], who had assumed $\theta$ = 40° based on the maximum slopes of the terrain in proximity of the sampling site). In line with Walsh & Ballouz et al. (2022), our result implies that the near-surface is approximately half as dense as the asteroid globally (1190 ± 18 kg/m³) [Barnouin et al. 2019]. Combined with the average bulk density for hummocky and angular Bennu particles — 1540 ± 100 kg/m³ and 1720 ± 70 kg/m³, respectively [Ryan et al. in press] — we determine that the near-surface macroporosity is approximately 43–66%. Using the average sample grain density of 2800 ± 100 kg/m³ [Ryan et al. in press], the total porosity of the near-surface is thus 68–80%.

Such large values for the macroporosity of the Bennu's surface, particularly with respect to the asteroid's global macroporosity, were hypothesized to be due to the percolation of finer particles into the interior of the asteroid through surface processes such as mass movement [Walsh & Ballouz et al. 2022]. Such a process may lead to the development of a near-surface fine particle layer [Bierhaus et al. 2024], based on the morphology of small crater floors and the particle size distribution during sampling [Lauretta et al. 2022]. Vernazza et al. [2012] suggested that high porosity surfaces may be a general feature of asteroids, as emission features in the mid-infrared domain (7–25 μm) of large main-belt asteroids can be reproduced in the laboratory by preparing high-porosity (>90%) meteorite and/or mineral powders. The formation of the near-surface (top few millimeters) structure of those large asteroids may not be wholly different than on Bennu. As larger asteroids may be able to retain finer particulates, their added cohesion may allow for the formation of even more porous structures that those observed on Bennu.

*5.3 Penetration Dynamics of Impact-Generated Fragments*

Equipped with a direct measurement of $\theta$ through physical characterization of the sample, we used low-speed impact force models to evaluate the fate of impact fragments generated from the disruption of boulders on the asteroid's surface. We used the force model developed in Ballouz et al. [2021] for low-speed impacts into regolith surfaces based on the results of laboratory and numerical experiments to evaluate the dynamics of a fragment with mass $M_f$:

$$M_f \dot{U}_f = -M_f g + 5.2 \tan(\theta)\, \rho_b A U_f^{4/3} + 8 \tan(\theta)(\rho_b \delta)^{1/2} \left(\frac{A}{\pi}\right) g|z| + \sigma_c A, \qquad (6)$$

where $g$ is the local gravity, $A$ is the cross-sectional area of the fragment, $U_f$ is the fragment's impact speed, $\dot{U}_f$ is the time derivative of $U_f$, $\delta$ is the fragment's density, $|z|$ is its depth in the regolith, and $\sigma_c$ is the compressive strength of the regolith. A regolith's resistance to penetration is composed of three non-gravity force terms shown in Eq. (6): a speed-dependent drag term (2nd term on the right-hand-side), a depth-dependent pressure term (3nd term on the right-hand-side),, and a cohesive term (4th term on the right-hand-side). On Bennu, the drag term dominates over the others; therefore, characterizing the velocity of ejecta fragments, in addition to the regolith geotechnical terms, is critical for evaluating the dynamics of penetration (or system ejection through bouncing). We used the simulation results presented in Sec. 4.2 to establish the initial conditions of ejecta fragments from disrupted boulders impacting regolith at low speeds. As shown in Fig. 19, for an impact disrupting a Bennu boulder, >96% of fragments that are ejected within 45° of the impactor's trajectory travel at speeds <1 m/s, with median values of 0.34 m/s.

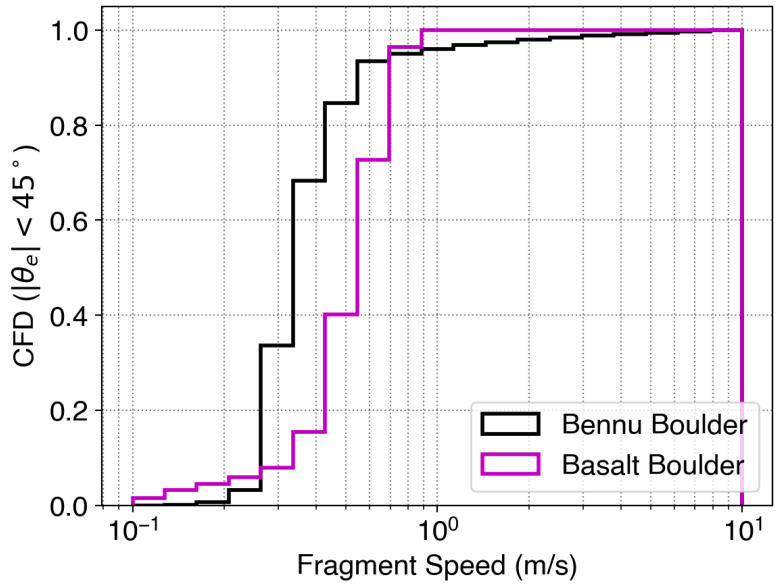

**Figure 19.** The cumulative frequency distribution (CFD) of fragments ejected from the simulated Bennu and basalt boulders at an angle, $\theta_e$, within 45° of the impactor's trajectory.

Considering spherical boulder fragments with diameters, $D$, of 4 and 40 m impacting Bennu regolith at speeds, $U_f$, of 0.3 and 2.0 m/s, we numerically integrated Eq. (6) using the Python package *scipy* and its *odeint* function, which solves a system of differential equations using the *lsoda* integrator [Virtanen et al. 2020], until the magnitude of the velocity reaches zero. For this calculation, we assumed spherical fragments, $g = 7 \times 10^{-5}$ m/s$^2$ [Barnouin et al. 2019], $\rho_b = 725$ kg/m$^3$ [Ryan et al., in press], $\theta = 32°$ (Sec. 5.1), and $\delta = 1650$ kg/m$^3$ [Ryan et al., in press]. The latter three quantities are constrained by the physical analysis of the Bennu samples. We assumed that the compressive strength term of the regolith is much smaller than the drag and pressure terms, as the cohesive bonding of Bennu particles is small [Jardine et al. 2025]. Figure 20 shows the evolution of the fragments' non-dimensionalized depth from the surface, $z/D$, and non-dimensionalized speed, $U_f/v_{esc}$, where $v_{esc}$ is Bennu's escape speed. We find that both fragment sizes penetrate approximately 0.3 and 0.65 boulder diameters into the regolith after impacting it at speeds at 0.3 and 2.0 m/s, respectively. The impact speeds considered are representative of the median and 99% percentile fragment speeds from Sec. 4.2. In both cases, the fragments are retained on the surface of the asteroid.

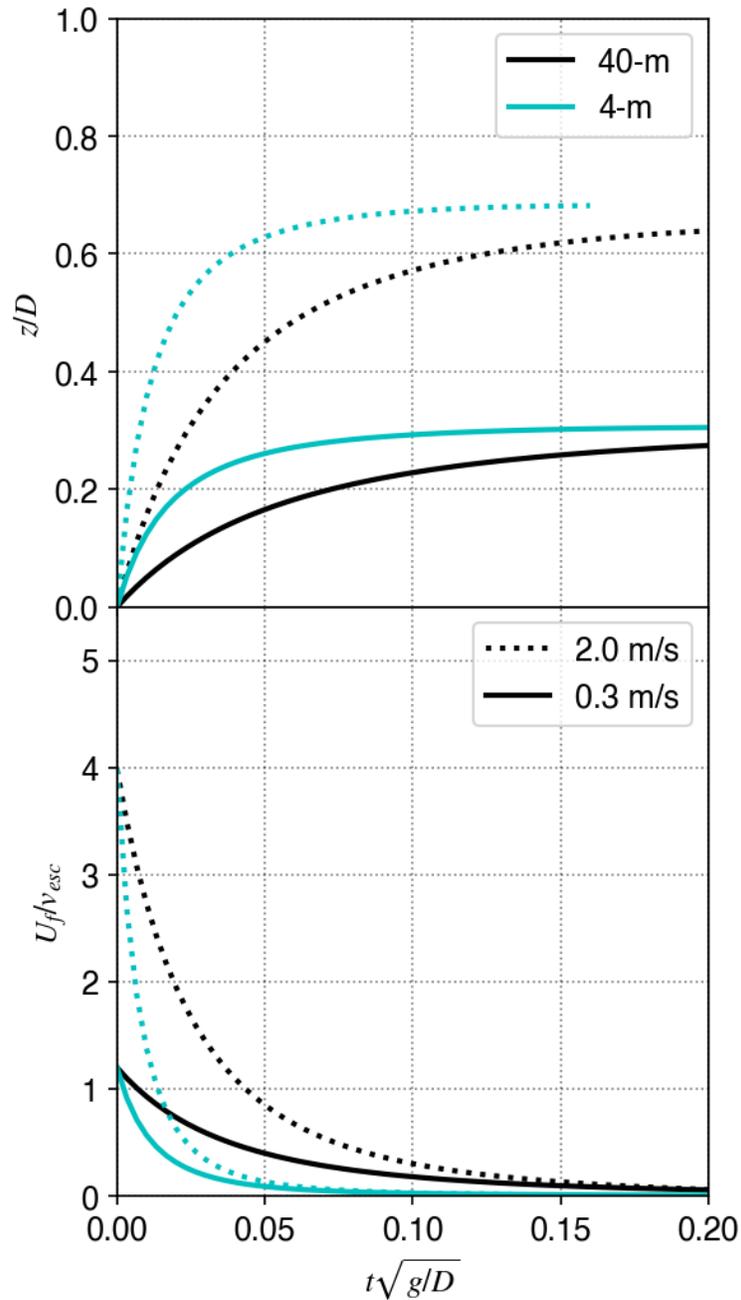

**Figure 20.** Dynamics of a 40-m-diameter (black) and a 4-m-diameter (cyan) fragment penetrating into Bennu regolith at 2.0 m/s (dotted curves) and 0.3 m/s (solid curves). **a,** non-dimensional depth, $z/D$, where $z$ is fragment position from the surface and $D$ is fragment diameter, as a function of non-dimensional time, $t$. **c,** evolution of fragment velocity, $U_f$, normalized by the Bennu's escape speed, $v_{esc}$.

# 6. Regolith Production and Retention on Kilometer-scale NEAs through Disruption

We have introduced a physical mechanism where disruptive impacts of particles on a small asteroid's surface can result in the retention of most of the collisional fragments' mass, because these fragments predominantly eject in the direction of the impactor (that is, toward the surface; e.g., Figure 16), and the weak, porous regolith arrests their momentum despite their high speed relative to the escape velocity of the asteroid. The Bennu samples offer direct evidence that collisions result in the retention of centimeter-scale impact targets. In this section, we assess the extent to which our proposed mechanism may operate on the surface of Bennu and small NEAs in general. We also present implications for understanding the recent history ($\lesssim$1 Myr) of the Bennu samples.

*6.1 Observational Evidence of Collisional Fragment Retention on the Surface of Bennu*

Figure 21 shows an example of a disrupted boulder on Bennu located at the center of a 128-m-diameter crater (21N,189E). The images show a cluster of boulders separated by through-going fractures. The appearance of this collection of boulders is reminiscent of "campfire" structures previously reported on the surface of Bennu [e.g., Walsh et al. 2019]. We used high-resolution 3D point cloud data acquired by the OSIRIS-REx Laser Altimeter (OLA) [Daly et al. 2017, Barnouin et al. 2019, Daly et al. 2020] to clarify the spatial relationship between the cluster of boulders.

The OLA data show that (i) there is a circular depression at the center of the boulder cluster, (ii) the linear features that separate the component boulders go through them, with some OLA returns in the gaps, and (iii) the linear features have a radial pattern with the center of the depression as an origin. We therefore infer that the cluster consists of the fragments of a disrupted boulder. The presence of a disrupted boulder in the center of such a large crater (relative to Bennu's size) suggests that the original boulder may have partially armored Bennu's surface against the formation of an even larger crater. It also indicates that fragments produced by larger cratering impacts can be retained on the surface through the mechanism we have proposed.

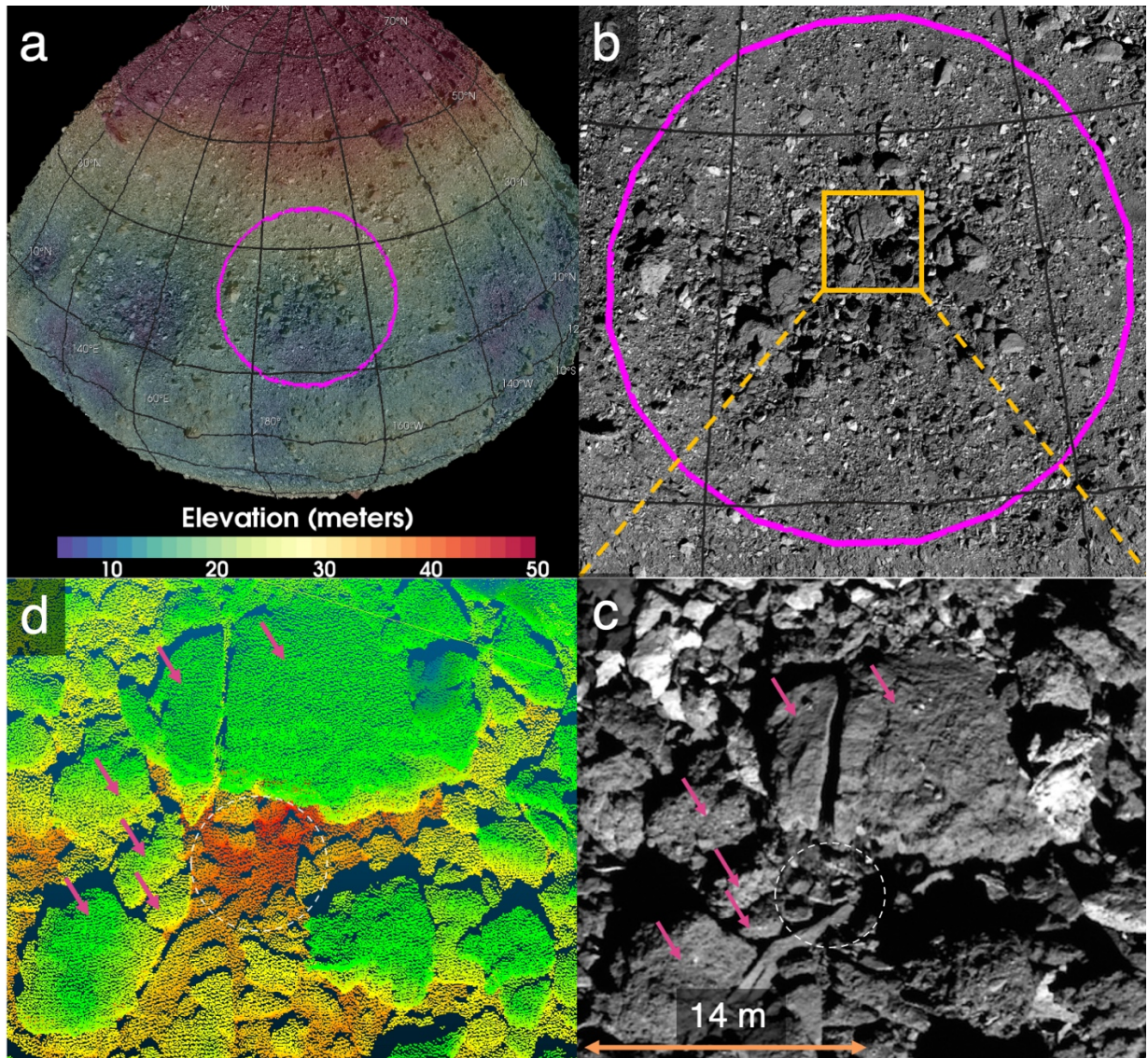

**Figure 21.** Example of a disrupted boulder on Bennu located at the center of a 128-m-diameter crater (21N,189E). **a,** The OLA v21 shape model of Bennu [Daly et al. 2021] with the global basemap projected onto the surface [Bennett et al. 2020]. The 128-m diameter crater is outlined in magenta. **b,** Close-up of the crater. The yellow square highlights the region of the disrupted boulder that is shown in more detail in **c**. **c,** PolyCam image of the disrupted boulder, identified by the radial concentric through-going fracture pattern. **d**, OLA DTM of **c**, revealing a central pit (red is the lowest point) at the point of origin of the radial fractures. Pink arrows in **c** and **d** highlight the same fragments.

*6.2 The Residence Time of Particles on the Surface of Bennu*
To clarify the extent to which disruptive impacts meaningfully contribute to the production of regolith, we calculated the residence time of surface particles on Bennu before they are catastrophically disrupted.

If an object's intrinsic strength controls its response to impacts, which is true for boulders and small coherent asteroids, then the catastrophic disruption threshold in the strength regime, $Q^*_S$, is given by

$$Q^*_S = \frac{1}{2}\frac{M_p U^2}{M_T} = q_S R_T^{-\mu_S} U^{2-3\mu_S} \quad (7)$$

where $q_S$ and $\mu_S$ are disruption scaling constants that depend on material properties. In Ballouz et al. [2020], the value of the exponent $\mu_S$ was obtained through observations of craters on boulders, but $q_S$ could not be directly determined as no assessment of this coefficient for coherent carbonaceous asteroids had been made. Using Eq. (7) and the results of our laboratory experiments, we estimated a value for $q_S = 1.48$, assuming $\mu_S = 0.47$ [Ballouz et al. 2020]. Thus, we suggest the following scaling relationships for evaluating the potential for disruption of impacts into solid Bennu-like targets:

$$Q^*_S = 1.48 R_T^{-0.47} U^{0.59} \quad (8)$$

Using Eq. (8), we calculated the diameter of a meteoroid that would have $Q = Q^*_S$ if impacting the surface of Bennu in the main belt or in near-Earth space. In the main belt, the average impact probability $P_i = 2.9 \times 10^{-18}$ km$^{-2}$, and the average impact speed $<U> = 5.3$ km/s [Bottke et al. 2005]. For near-Earth space, we considered the impact flux given by Brown et al. [2002] and Grün et al. [1985] for $<U> = 18.5$ km/s. $D_{imp}$ was then determined for the main belt and near-Earth space using our derived value of $Q^*_S$ for $U = <U>$ in each environment.

We used the approach of Ballouz et al. [2020] to determine the total population of potential disruptive colliders. As described therein, we first considered the CSFD, $N_{C, MBA}$, of main belt asteroids (MBAs) calculated from observations of the MBA size distribution and models of their collisional evolution [2]. By numerically differentiating $N_{C, MBA}$, we derived the incremental size frequency distribution $N_{I, MBA}$ of MBAs with diameters $D$:

$$N_{I, MBA}(D_k) = N_{C, MBA}(\geq D_k) - N_{C, MBA}(\geq D_{k+1}) \quad (9)$$

where $k$ is the index of the logarithmically binned CSFD data and $D_{imp,k+1} > D_{imp,k}$. Then, the number of disruptive impacts, $N_{MBA}$, over a mean time interval, $t_{coll}$, of a boulder with diameter $D_B$ by an object with diameter $D_{imp}$ is given by

$$N_{MBA} = N_{I, MBA}(D_{imp}) \times P_i \times \frac{1}{2}\left(\frac{D_{imp}}{2} + \frac{D_B}{2}\right)^2 \times t_{coll}, \quad (10)$$

where $D_B$ is the boulder diameter and $D_B = 2 R_T$. The third term on the right side of Eq. (10) is the collisional cross-section divided by 2, as we approximate that a boulder resting on the surface of an asteroid is shielded from half of all potential impactors. The mean collisional lifetime was then calculated by setting $N_{MBA} = 1$ and solving for $t_{coll}$.

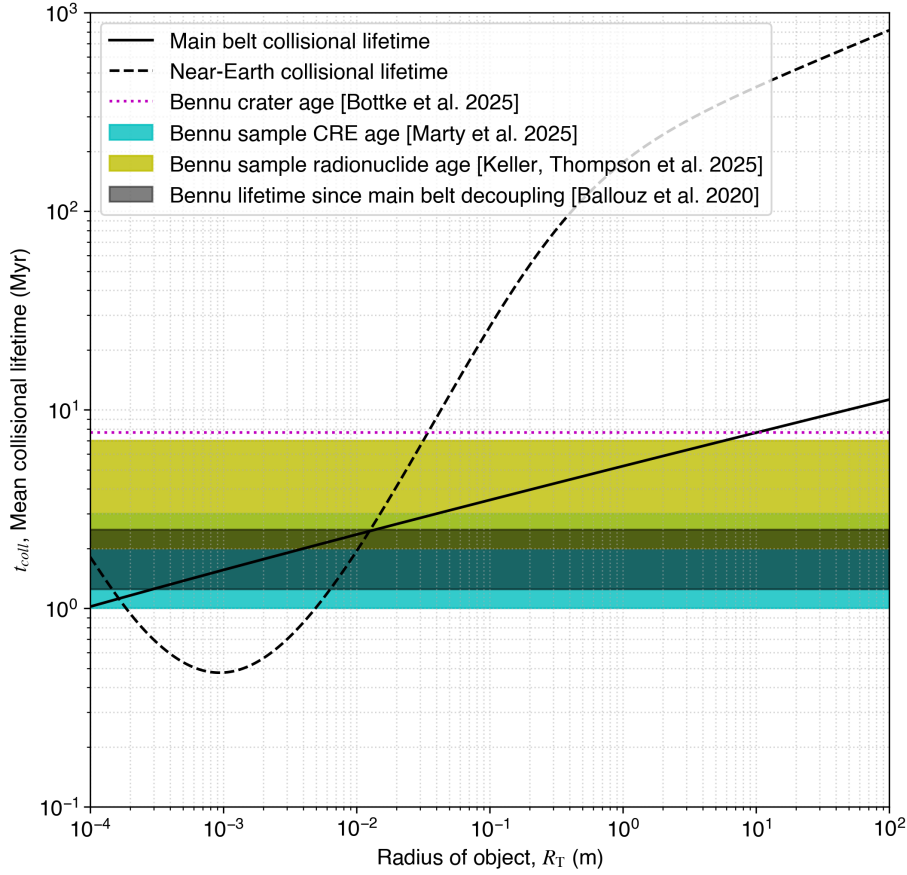

**Figure 22.** The mean collisional lifetime of Bennu (and Bennu-like) particles, with radii from 0.1 mm to 100 m, on the surface of an asteroid as a function of their size, assuming impacts from main belt (solid black curve) and near-Earth cometary and asteroidal sources (dashed black curve). Relevant ages for the surface of Bennu from crater chronology and radiometric dating of samples are also shown. The dotted pink horizontal line marks the median crater surface age of 7.7 Myr for Bennu [Bottke et al. 2025]. The range of cosmic-ray exposure (CRE) ages (1–3 Myr) and radionuclide ages (2–7 Myr) determined for Bennu samples are shown in the cyan- and yellow-shaded regions [Marty et al. 2025, Keller & Thompson et al. 2025]. Bennu's lifetime since it decoupled from the main belt (1.25–2.5 Myr) is shown in the grey-shaded region [Ballouz et al. 2020].

For near-Earth space, we performed a similar analysis by using the cumulative impact flux from two sources: cometary dust [Grün et al. 1985] (see their equation (A3)), which dominates at smaller sizes ($\lesssim 2$ cm), and meteoroids from asteroidal sources ($\gtrsim 2$ cm) [Brown et al. 2002] (see our Equation 3). We normalized the cumulative impactor flux to the cross-sectional area of a surface boulder, divided by 2, as above. Then, the number of disruptive impacts in near-Earth space, $N_{NEA}$, over $t_{\text{coll}}$ was determined by numerically integrating the flux equations, such that

$$N_{NEA} = N_{\text{I,NEA}}(D_{\text{imp}}) \times t_{\text{coll}} \tag{11}$$

The mean collisional lifetime in near-Earth space was then calculated by setting $N_{NEA} = 1$, and solving for $t_{coll}$.

Fig. 22 shows the calculated $t_{coll}$ for Bennu (and Bennu-like) particles, as a function of their size, on the surface of an asteroid residing in the main belt (solid black curve) and in near-Earth space (dashed black curve). Ages of the surface of Bennu from crater chronology and radiometric dating of samples are also shown.

Our calculation shows that, if Bennu resided in the main belt for ~7.7 Myr [Bottke et al. 2025], as indicated by large craters on its surface, boulders of up to 20 m diameter would have been disrupted. Therefore, surface boulders with diameters ≲ 20 m are either the products of a disruption event or only recently excavated by mechanisms such as mass wasting [e.g., Jawin et al. 2020]. Of the >3000 Bennu boulders with diameters >1 m, only a few tens have diameters > 20 m [DellaGiustina & Emery et al. 2019]. Therefore, only a small minority of individual boulders on Bennu's surface may date back to its formation as a rubble-pile. Thus, the vast majority of boulders on Bennu could be the result of collisional disruption events that took place after Bennu formed, or could be recently excavated by impacts that form large craters or mass movement events. The rate of boulder disruption decreases significantly once Bennu is no longer coupled to the main belt. That allows for the accumulation of craters on boulders in timescales ≳100 Myr.

In near-Earth space, $t_{coll}$ is controlled by cometary sources, which dominate over asteroidal sources for the disruption of objects ≲1 m diameter. Based on the various chronometers we considered, these impacts are relevant to the disruption of objects ≲8 cm over Bennu's lifetime. Critically, $t_{coll}$ is shorter in near-Earth space than in the main belt for objects with radii between ~0.01 and 1 cm, ranging from ~0.5 to 2.5 Myr and reaching a minimum for 1-mm-radius objects. The particle SFD of the Bennu sample collection shows a turnover (i.e., depletion) of particles at this size [Ryan et al. in press]. Our results suggest that the depletion of millimeter-sized objects may result from their degradation by impacts, potentially contributing to the overall scarcity of dust on the surface [Hamilton et al. 2019, Rozitis et al. 2020].

The relatively short $t_{coll}$ in near-Earth space indicates that collisions play an important role in understanding the recent evolutionary history of the Bennu samples, as the particles have sizes that range from a few tens of microns up to a few centimeters [Lauretta & Connolly et al. 2024, Ryan et al. in press]. Marty et al. [2025] found CRE ages, which pertain to the exposure of material in the top ~1 m of the surface, of 1–3 Myr for the Bennu samples they analyzed. Keller & Thompson, et al. [2025] showed that although the cosmogenic noble gases and radionuclide CRE ages of particles collected directly from the surface by the TAGSAM contact pads range from 2 to 7 Myr, the density of solar energetic particle tracks and craters on the particles suggest shorter direct exposure to space (<85 Kyr). The older CRE ages are concurrent with Bennu's near-Earth lifetime calculated based on crater chronology on meter-scale Bennu boulders [Ballouz et al. 2020]. Our work confirms that the younger age [Keller & Thompson, et al. 2025] is a result of recent exposure to space: those particles, or their immediate precursors, were probably not disrupted, because $t_{coll}$ > 0.5 Myr for all particles collected by OSIRIS-REx. Rather, the younger age likely tracks the most recent excavation of material through mass movement or the formation of Hokioi crater (the site of OSIRIS-REx's sampling) [Jawin et al. 2020, Barnouin et al. 2022].

The accumulation of regolith on small rubble-pile asteroids from the disruption of particles and boulders is balanced by their loss to space from larger impacts that form craters in the regolith in the gravity regime. Takaki et al. [2022] showed that the upper 2-4 m of regolith on Ryugu is excavated on <1 Myr timescales based on gravity scaling. However, analysis of returned Ryugu samples show that their CRE ages are much older: 5 Myr [Okazaki et al. 2022]. This discrepancy suggests that the rate of regolith loss by impact cratering may be slower than assumed by gravity-scaled cratering models. Our proposed model of regolith generation from impacts may counterbalance this loss. More detailed modeling of the sources and sinks of regolith on small rubble piles is required to fully understand the regolith renewal and evolution process.

*6.3 Micrometeoroid Impacts as the Source of Bennu's Particle Ejection Events*

The OSIRIS-REx spacecraft observed episodes of particles ejecting from the asteroid's surface [Lauretta & Hergenrother et al. 2019, Hergenrother et al. 2020]. One hypothesis for the driver of this observed activity was micrometeoroid bombardment [Bottke et al. 2020]. This impactor population would likely be the same as that which formed the craters we observe on the Bennu samples. In Fig. 3b, we highlight a crater surrounded by a circular spall region and a larger, extended spall region demarcated by what appears to be the original pre-impact surface at the particle's edge. The extended spall region is approximately 5.5 mm long and wide. If an impact onto this particle ejected a spalled fragment, that fragment would have an axis ratio of ~0.16. In comparison, particles that were observed ejecting from the surface of Bennu had effective diameters of 2.2 to 61 mm (median of 7.4 mm) and axis ratios of 0.07 to 1.0 (median of 0.27) [Chesley et al. 2020], corresponding to a platy shape.

The Bennu sample also contained platy angular centimeter-scale particles that are able to fit together like puzzle pieces [Ryan et al. 2025, in press]. Furthermore, the splitting of the angular stone OREX-800055-0 led to the production of platy and bladed particles like those observed in ejection events, albeit at smaller scales [Ryan et al. 2025, in press]. Taken together, these observations indicate that the angular morphology may be the source of many of the ejected particles, and both low and high specific energy processes can produce platy fragments. Thus, this impact feature observed on a sample supports the hypothesis that micrometeoroid bombardment is responsible, at least in part, for the observed ejection of platy particles off Bennu's surface [Bottke et al. 2020].

As described in Sec. 6.2, most craters on the Bennu samples likely formed by cometary dust particles. However, the morphological properties of some craters observed on the samples suggest that asteroidal sources may have contributed as well, as had been suggested by Ballouz et al. [2020] for the formation of decimeter- and larger-sized craters on Bennu's boulders. In particular, the unique morphology of crater 2 (Fig. 5 in Section 2) suggests that the Bennu particles were exposed to a population of dense impactors. The ratio of impactor penetration depth to density developed by Okamoto et al. [2015] suggests that the impactor that formed crater 2 was over-dense by a factor of 1.58 compared to the average impactor. Assuming that the average impactor is a cometary particle with a density of 0.9 g/cm$^3$, crater 2's impactor would have had a density of 1.75 g/cm$^3$, which is at the high end of the bulk densities measured for Bennu samples, thus far, and within the range of CI carbonaceous chondrites [Ryan et al. in press]. This suggests that Bennu also experienced impacts from material originating from other carbonaceous asteroids.

## 7. Summary and Outlook

An open question in small body geophysics has been the origin of NEA regolith. Is regolith fully inherited from parent bodies, or can it be produced by in situ processes, such as impact, despite their microgravity environments? Here, through a combination of sample analysis, laboratory experiments, and numerical simulations, we have shown that the majority of fragments produced by disruption of Bennu's boulders are retained on the surface of the asteroid.

We summarize our main findings and their implications for the physical properties of and surface evolution of Bennu and similar NEAs:

Major findings:
- For most Bennu stones that we examined, multiple faces exhibit cratering. These observations imply that the surface of Bennu was mobilized in the recent past, and that the asteroid's near-surface has experienced regolith gardening and/or mass movement, in agreement with inferences from remote observations [Jawin et al. 2020, Barnouin et al. 2022]. This is also consistent with the younger exposure age estimated for contact pad samples based on solar energetic particle track and crater densities in comparison to their CRE ages [Keller & Thompson, et al. 2025].

- Boulders on Bennu with diameters ≲ 20 m may be the products of catastrophic disruption of larger boulders. Our findings indicate that these collisional products were retained on the asteroid because they were able to penetrate Bennu's weak surface, which dissipated the excess energy that would otherwise cause them to escape by rebounding off the surface.

- The collisional lifetime of particles on the surface of Bennu, $t_{coll}$, is at a minimum for 1-mm-radius objects. The particle SFD of the returned Bennu samples also shows a turnover (i.e., depletion) at particles of this size and smaller [Ryan et al. in press]. The depletion of millimeter-sized objects may be a result of impact degradation of particles of those sizes, and the decreased production of fine particles through compaction of surface from cratering impacts. These effects may contribute to the observed scarcity of dust on the surface of Bennu [Hamilton et al. 2019, Rozitis et al. 2020].

Sample analysis:
- The presence of craters on Bennu stones provides direct evidence that the OSIRIS-REx mission, through preparation and care [Lunning et al. 2025], was able to preserve the physical context of Bennu regolith from the moment of sampling through transport to the NASA JSC curation facility.

- The range of crater diameters and depths on stones (median $d/D = 0.36 \pm 0.1$) overlaps with that found for boulders on Bennu (mean $d/D = 0.33 \pm 0.08$ [Ballouz et al. 2020]). This similarity suggests the likelihood of structural similarity between the stones and boulders, as they appear to respond in a self-similar way to impacts. Thus, the porosities observed in the samples may be reflective of the total porosities of Bennu's boulders.

- Some craters on stones have morphologies different than the typical central pit and spall features. Crater 2 (Fig. 5c,d) lacks a clear spall region and has a deep triangular pyramidal shape with a bulbous pit, indicating an impactor with a Bennu- and CI-like density. This suggests that the sampled regolith experienced impacts sourced from other carbonaceous asteroids.

- Extended spall features surrounding some craters on stones support the hypothesis that micrometeoroid bombardment may be responsible, at least in part, for the observed ejection of platy particles off Bennu's surface [Bottke et al. 2020].

- Computer-vision analysis of videography of the pour-out of the sampled Bennu material indicates an angle of friction of 31.8° ± 2.1°, consistent with estimates from remote sensing [Barnouin et al. 2022, Robin et al. 2024]. Using this updated value with results from the reconstruction of the OSIRIS-REx sampling event [Walsh & Ballouz et al. 2022] gives a near-surface bulk density, $\rho_b$ = 580–870 kg/m$^3$. Using the average sample grain density of 2800 ± 100 kg/m$^3$ [Ryan et al. in press], the total porosity of the near surface of Hokioi crater on Bennu is 68–80%. Further analysis of OSIRIS-REx remote sensing data may provide insight into whether this finding applies more generally to Bennu's whole surface.

Impact experiments:

- Impact experiments into simulant show that the CSFD power-law exponent, $s$, for fragments produced by catastrophic disruption agrees with that of the OSIRIS-REx sampling site and the sample itself [Burke et al 2021, Ryan et al. in press], supporting the importance of disruptive collisions in the evolution of Bennu's regolith.

- The impact experiments also show that, for the simulant studied here with porosities up to 45%, the disruption threshold $Q^*_S$ = 724 J/kg, which is substantially smaller than that reported for some ordinary chondrite meteorites and Allende. This value of $Q^*_S$ is closer in value to experimental findings for carbonaceous chondrites with lower porosity than the simulant, specifically Aguas Zarcas (CM2) and NWA 4502 (CV3) [e.g., Flynn et al. 2018, Flynn et al. 2025]. This means that the CI simulant is significantly less resistant to impact disruption than ordinary chondrites.

- The post-impact XCT scan of simulant targets reveals that the material in the vicinity of the impact region was compacted, supporting the hypothesis of Cambioni et al. [2021] that the production of fines on Bennu is suppressed due to compaction cratering.

Numerical models:

- We calibrated a SPH code to simulate impacts into a Bennu boulder and find that the best match to the experimental results for the Weibull parameters are $m$ = 9.5 and $k$ = 5 x 10$^{41}$ cm$^3$. We use these values to estimate a compressive strength for the material of $\sigma_c$ = 0.33–0.62 MPa for a 1-m-diameter Bennu boulder, which is within the range estimated for Bennu's boulders based on remote characterization of craters (0.44–1.7 MPa) [Ballouz et al. 2020].

Remote sensing:

- Our analysis of OSIRIS-REx PolyCam and OLA data revealed evidence of the disruption of a 10-m boulder on the surface Bennu. These observations provide supporting evidence of the in situ collisional processing of a large boulder into smaller particles that are retained on the micro-gravity surface of a 500-m diameter NEA. The numerical simulations indicate that such impacts would result in the majority of the fragmented boulder mass being directed into the asteroid surface at speeds < 1 m/s.

As we have shown here, analysis of the millimeter-scale crater population on Bennu samples provides crucial information on the history and physical nature of this carbonaceous material. Additional analysis of the physical and thermal properties of other returned extraterrestrial samples would allow us insight into regolith evolution other Solar System bodies. Further physical and thermal analysis will allow us to further test and refine the regolith development framework we have presented here.

**Acknowledgements**

This work is supported by NASA under Contract NNM10AA11C issued through the New Frontiers Program. We are grateful to the entire OSIRIS-REx Team for enabling the return and analysis of samples from Bennu. We thank the Astromaterials Acquisition and Curation Office, part of the Astromaterials Research and Exploration Science (ARES) Division at Johnson Space Center, for their efforts in recovery, preliminary examination, and long-term curation of the Bennu samples. We also greatly appreciate support from the OSIRIS-REx Sample Analysis Micro-Information System (SAMIS) Team.  R.-L.B. and O.S.B. are also supported by NASA New Frontiers Data Analysis Program grant number 80NSSC22K1035. R.-L.B. additionally acknowledges support from the JHU/APL Sabbatical Fellowship. We thank the developers of *miluphCUDA* for providing access to their code. We thank Christopher DeBuhr for assistance in acquiring SEM images of sample OREX-800123-0. We thank the two anonymous referees for providing constructive reviews of the original submitted manuscript.

**Data Availability**

The instrument data products underlying the findings of this work will be available on AstroMat.org via the DOIs in Table A2. Measurement, experimental, and simulation data are presented in the manuscript and the appendix. *miluphCUDA* is an open-source software that is freely distributed by its developers.

# Appendix

## A.1 Impact Experiment Setup

The HyFIRE facility includes a two-stage gas gun, a vacuum chamber, and a suite of diagnostic tools which include highspeed cameras capable of capturing the post-impact dynamics of fragments as a target is disrupted. The impact chamber was pumped down to 5 Torr. We used front and side cameras to capture the fracturing of the targets at 200,000 to 2 million frames per second, and a top-view camera at 20,000 frames per second, to capture the post-disruption dynamics. Calibration images were taken from each camera point of view before each experiment. The cube targets were suspended inside the vacuum chamber using fishing wire to mitigate against the influence of surface contact on post-impact velocities of fragments. Projectiles made of an aluminum 2017 alloy and 440 stainless steel were fired at the center of the targets (0° impact angle) at impact speeds of 5.3–5.4 km/s (Table 1), the average speed of collisions between asteroids in the main belt [e.g., Bottke et al. 2005]. The focus of these experiments was to assess the catastrophic disruption threshold of the simulant and the three-dimensional velocity distribution of fragments following impact.

## A.2 SPH Model

Our model incorporates the Tillotson EOS [Tillotson 1962, Melosh 1989]; the material damage model derived by Grady & Kipp [1980] and first implemented by Benz & Asphaug [1994]; and a p-alpha porosity model [Jutzi et al. 2010]. Table 2 summarizes the Tillotson EOS and p-alpha model parameters used in our simulations and each value's associated reference. We used the grain density, $\rho_g$, of the simulant that we measured in our experimental calibration runs. For the simulations of impacts on to Bennu boulders, we used the grain density estimated from the bulk mineralogy of a homogenized Bennu sample (sample ID) obtained through XRD [Ryan et al., in press]. The Tillotson parameters are similar to those that Nakamura et al. [2022] used for their 2D impact simulations into the Ryugu parent body which they termed the "soft" case. We use a shear modulus and bulk modulus that were estimated from seismic wave measurements of an angular centimeter-scale Bennu sample, OREX-800123-0 [Hildebrand et al. personal communication, 2025]. We use p-alpha parameters derived for pumice [Jutzi et al. 2010] for values that have not yet been determined for Bennu or Bennu-like materials.

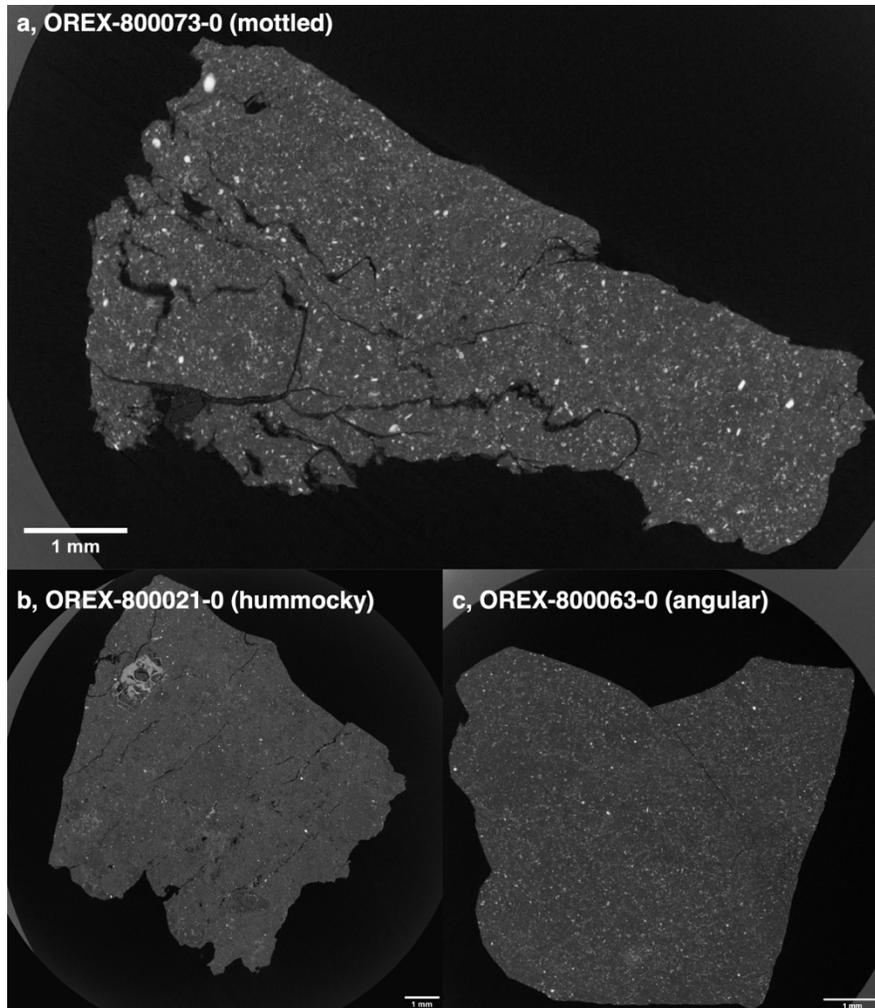

Figure A1. Examples of the XCT cross-sections of Bennu samples with impact craters that shows the range in internal fracture density observed in the samples. a, XCT cross-section of the mottled stone OREX-800073-0. b, XCT cross-section of the hummocky stone OREX-800021-0. c, XCT cross-section of the angular stone OREX-800063-0.

| Sample ID | Classification | Shape Model | Craters | $N_{craters}$ |
|---|---|---|---|---|
| OREX-800014-0 | Mottled | SLS | Y | 14 |
| OREX-800016-0 | Unclassified | SLS | Y | 4 |
| OREX-800018-0 | Unclassified | SLS | Y | 17 |
| OREX-800021-0 | Hummocky | XCT | Y | 16 |
| OREX-800047-0 | Unclassified | XCT | Y | 11 |
| OREX-800055-0 | Angular | XCT | Y | 4 |
| OREX-800063-0 | Angular | XCT | Y | 6 |
| OREX-800067-0 | Angular | XCT | Y | 7 |
| OREX-800073-0 | Mottled | XCT | Y | 3 |
| OREX-800113-0 | Hummocky | XCT | Y | 4 |

| Sample ID | Classification | Shape Model | Craters | $N_{craters}$ |
|---|---|---|---|---|
| OREX-800122-0 | Hummocky | XCT | Y | 6 |
| OREX-800123-0 | Angular | XCT | Y | 15 |
| OREX-800129-0 | Angular | XCT | Y | 11 |
| OREX-800097-0 | Angular | XCT | Y | 19 |
| OREX-800098-0 | Angular | XCT | Y | 6 |
| OREX-800019-0 | Angular | XCT | N | - |
| OREX-800089-0 | Hummocky | XCT | N | - |
| OREX-800026-0 | Hummocky | SLS | N | - |
| OREX-800017-0 | Angular | SLS | N | - |
| OREX-800096-0 | Unclassified | XCT | N | - |
| OREX-800100-0 | Mottled | XCT | N | - |
| OREX-800087-0 | Hummocky | SLS | M | - |
| OREX-800088-0 | Hummocky | XCT | M | - |
| OREX-800118-0 | Hummocky | XCT | M | - |
| OREX-800020-0 | Angular | SLS | M | - |
| OREX-800023-0 | Mottled | XCT | M | - |
| OREX-800027-0 | Unclassified | XCT | M | - |
| OREX-800099-0 | Hummocky | XCT | M | - |
| OREX-800134-0 | Angular | XCT | N | - |

**Table A1.** Summary of crater mapping on Bennu stones. The Classification column specifies the type of stone, as defined in Lauretta & Connolly et al. (2024) and Connolly & Lauretta et al. (2025)). The Shape Model column refers to the model's source data (SLS, structured light scanning; XCT, X-ray computed tomography). The Craters column describes our assessment of whether craters are present (Y, yes; N, no; M, maybe). $N_{craters}$ is the number of craters mapped on each stone.

| Sample ID | Laboratory | Analysis | DOI |
|---|---|---|---|
| OREX-800014-0 | Johnson Space Center | SLS | 10.60707/64vm-zd18 |
| OREX-800016-0 | Johnson Space Center | SLS | 10.60707/w2w2-0448 |
| OREX-800017-0 | Johnson Space Center | SLS | 10.60707/fymn-n257 |
| OREX-800018-0 | Johnson Space Center | SLS | 10.60707/pnfj-nc59 |
| OREX-800019-0 | Johnson Space Center | XCT | 10.60707/nxge-6x60 |
| OREX-800019-0 | Johnson Space Center | SLS | 10.60707/46vk-8e09 |
| OREX-800020-0 | Johnson Space Center | SLS | 10.60707/c99t-qs13 |
| OREX-800021-0 | Johnson Space Center | XCT | 10.60707/czsb-b064 |

| Sample | Laboratory | Technique | DOI |
|---|---|---|---|
| | Johnson Space Center | SLS | 10.60707/s4m1-4b95 |
| OREX-800023-0 | Johnson Space Center | XCT | 10.60707/nptt-2x39 |
| | Johnson Space Center | SLS | 10.60707/3cqf-cm22 |
| OREX-800027-0 | Johnson Space Center | XCT | 10.60707/140w-3640 |
| OREX-800047-0 | Johnson Space Center | XCT | 10.60707/mjvq-aa31 |
| OREX-800049-0 | Johnson Space Center | XCT | 10.60707/crnr-cf55 |
| OREX-800054-0 | Johnson Space Center | XCT | 10.60707/6y35-r522 |
| OREX-800055-0 | Johnson Space Center | XCT | 10.60707/2c1c-dz58 10.60707/8pbm-t622 |
| | Johnson Space Center | SLS | 10.60707/9tad-kq08 |
| OREX-800067-0 | Johnson Space Center | XCT | 10.60707/hvs8-a166 |
| | Johnson Space Center | SLS | 10.60707/dc7d-1949 |
| OREX-800073-0 | Johnson Space Center | XCT | 10.60707/brk7-px44 |
| OREX-800088-0 | Johnson Space Center | XCT | 10.60707/8w70-4b88 10.60707/e4ys-8z20 |
| | Johnson Space Center | SLS | 10.60707/c2ga-7937 |
| OREX-800088-5 | Johnson Space Center | XCT | 10.60707/e4ys-8z20 |
| OREX-800089-0 | Johnson Space Center | XCT | 10.60707/3ds1-5396 |
| | Johnson Space Center | SLS | 10.60707/a3qn-ak31 |
| OREX-800096-0 | Johnson Space Center | XCT | 10.60707/5dz0-7d81 |
| OREX-800097-0 | Johnson Space Center | XCT | 10.60707/z0gq-5c10 |
| OREX-800098-0 | Johnson Space Center | XCT | 10.60707/zvgn-8178 |
| OREX-800099-0 | Johnson Space Center | XCT | 10.60707/5kkn-q343 |
| OREX-800107-0 | Johnson Space Center | XCT | 10.60707/jnv8-1e46 |
| OREX-800117-0 | Johnson Space Center | XCT | 10.60707/zsgj-ed85 |
| OREX-800118-0 | Johnson Space Center | XCT | 10.60707/xe71-ec20 |
| OREX-800123-0 | Johnson Space Center | XCT | 10.60707/5m56-kv71 |
| | University of Calgary | SEM | 10.60707/gh6s-0e97 |
| OREX-800129-0 | Johnson Space Center | XCT | 10.60707/dnfj-mf92 |
| OREX-800134-0 | Johnson Space Center | XCT | 10.60707/z4yp-y044 |
| Bulk Sample in TAGSAM | Johnson Space Center | ARMCam | 10.60707/3p64-9472 |

**Table A2.** Samples studied in this work, the laboratories where they were analyzed, the analytical techniques applied, and the DOIs of the resulting datasets.